\documentclass[english,journal,12pt,onecolumn,draftclsnofoot]{IEEEtran}
\usepackage[T1]{fontenc}
\usepackage{geometry}
\usepackage{units}
\usepackage{amsmath}
\usepackage{amsthm}
\usepackage{amssymb}
\usepackage{graphicx}
\usepackage{cite}
\usepackage{bm}
\usepackage{graphicx}
\usepackage{url}
\usepackage{color}
\usepackage{algorithm}
\usepackage{algorithmic}
\usepackage{mathtools,array}

\makeatletter
\theoremstyle{plain}
\newtheorem{thm}{\protect\theoremname}

\usepackage{babel}
\usepackage[caption=false,font=footnotesize]{subfig}
\usepackage{balance}

\usepackage{tikz}
\usepackage{pgfplots}
\usepgfplotslibrary{groupplots}
\usepackage{graphicx}
\pgfplotsset{compat=newest}
\usepackage{units}
\usepackage{amsmath, amsbsy, amssymb}
\usepackage{tikzscale}
\usepackage{color}
\usetikzlibrary{plotmarks}
\usetikzlibrary{dsp,chains}
\usetikzlibrary{spy}
\usetikzlibrary{positioning,chains,fit,shapes,calc}
\usetikzlibrary{external}
\pgfplotsset{
colormap={jetlight}{rgb = (  1.00000000,   1.00000000,   1.00000000),rgb = (  0.99607843,   0.99607843,   0.99816176),rgb = (  0.99215686,   0.99215686,   0.99644608),rgb = (  0.98823529,   0.98823529,   0.99485294),rgb = (  0.98431373,   0.98431373,   0.99338235),rgb = (  0.98039216,   0.98039216,   0.99203431),rgb = (  0.97647059,   0.97647059,   0.99080882),rgb = (  0.97254902,   0.97254902,   0.98970588),rgb = (  0.96862745,   0.96862745,   0.98872549),rgb = (  0.96470588,   0.96470588,   0.98786765),rgb = (  0.96078431,   0.96078431,   0.98713235),rgb = (  0.95686275,   0.95686275,   0.98651961),rgb = (  0.95294118,   0.95294118,   0.98602941),rgb = (  0.94901961,   0.94901961,   0.98566176),rgb = (  0.94509804,   0.94509804,   0.98541667),rgb = (  0.94117647,   0.94117647,   0.98529412),rgb = (  0.93725490,   0.93725490,   0.98529412),rgb = (  0.93333333,   0.93333333,   0.98541667),rgb = (  0.92941176,   0.92941176,   0.98566176),rgb = (  0.92549020,   0.92549020,   0.98602941),rgb = (  0.92156863,   0.92156863,   0.98651961),rgb = (  0.91764706,   0.91764706,   0.98713235),rgb = (  0.91372549,   0.91372549,   0.98786765),rgb = (  0.90980392,   0.90980392,   0.98872549),rgb = (  0.90588235,   0.90588235,   0.98970588),rgb = (  0.90196078,   0.90196078,   0.99080882),rgb = (  0.89803922,   0.89803922,   0.99203431),rgb = (  0.89411765,   0.89411765,   0.99338235),rgb = (  0.89019608,   0.89019608,   0.99485294),rgb = (  0.88627451,   0.88627451,   0.99644608),rgb = (  0.88235294,   0.88235294,   0.99816176),rgb = (  0.87843137,   0.87843137,   1.00000000),rgb = (  0.87450980,   0.87647059,   1.00000000),rgb = (  0.87058824,   0.87463235,   1.00000000),rgb = (  0.86666667,   0.87291667,   1.00000000),rgb = (  0.86274510,   0.87132353,   1.00000000),rgb = (  0.85882353,   0.86985294,   1.00000000),rgb = (  0.85490196,   0.86850490,   1.00000000),rgb = (  0.85098039,   0.86727941,   1.00000000),rgb = (  0.84705882,   0.86617647,   1.00000000),rgb = (  0.84313725,   0.86519608,   1.00000000),rgb = (  0.83921569,   0.86433824,   1.00000000),rgb = (  0.83529412,   0.86360294,   1.00000000),rgb = (  0.83137255,   0.86299020,   1.00000000),rgb = (  0.82745098,   0.86250000,   1.00000000),rgb = (  0.82352941,   0.86213235,   1.00000000),rgb = (  0.81960784,   0.86188725,   1.00000000),rgb = (  0.81568627,   0.86176471,   1.00000000),rgb = (  0.81176471,   0.86176471,   1.00000000),rgb = (  0.80784314,   0.86188725,   1.00000000),rgb = (  0.80392157,   0.86213235,   1.00000000),rgb = (  0.80000000,   0.86250000,   1.00000000),rgb = (  0.79607843,   0.86299020,   1.00000000),rgb = (  0.79215686,   0.86360294,   1.00000000),rgb = (  0.78823529,   0.86433824,   1.00000000),rgb = (  0.78431373,   0.86519608,   1.00000000),rgb = (  0.78039216,   0.86617647,   1.00000000),rgb = (  0.77647059,   0.86727941,   1.00000000),rgb = (  0.77254902,   0.86850490,   1.00000000),rgb = (  0.76862745,   0.86985294,   1.00000000),rgb = (  0.76470588,   0.87132353,   1.00000000),rgb = (  0.76078431,   0.87291667,   1.00000000),rgb = (  0.75686275,   0.87463235,   1.00000000),rgb = (  0.75294118,   0.87647059,   1.00000000),rgb = (  0.74901961,   0.87843137,   1.00000000),rgb = (  0.74509804,   0.88051471,   1.00000000),rgb = (  0.74117647,   0.88272059,   1.00000000),rgb = (  0.73725490,   0.88504902,   1.00000000),rgb = (  0.73333333,   0.88750000,   1.00000000),rgb = (  0.72941176,   0.89007353,   1.00000000),rgb = (  0.72549020,   0.89276961,   1.00000000),rgb = (  0.72156863,   0.89558824,   1.00000000),rgb = (  0.71764706,   0.89852941,   1.00000000),rgb = (  0.71372549,   0.90159314,   1.00000000),rgb = (  0.70980392,   0.90477941,   1.00000000),rgb = (  0.70588235,   0.90808824,   1.00000000),rgb = (  0.70196078,   0.91151961,   1.00000000),rgb = (  0.69803922,   0.91507353,   1.00000000),rgb = (  0.69411765,   0.91875000,   1.00000000),rgb = (  0.69019608,   0.92254902,   1.00000000),rgb = (  0.68627451,   0.92647059,   1.00000000),rgb = (  0.68235294,   0.93051471,   1.00000000),rgb = (  0.67843137,   0.93468137,   1.00000000),rgb = (  0.67450980,   0.93897059,   1.00000000),rgb = (  0.67058824,   0.94338235,   1.00000000),rgb = (  0.66666667,   0.94791667,   1.00000000),rgb = (  0.66274510,   0.95257353,   1.00000000),rgb = (  0.65882353,   0.95735294,   1.00000000),rgb = (  0.65490196,   0.96225490,   1.00000000),rgb = (  0.65098039,   0.96727941,   1.00000000),rgb = (  0.64705882,   0.97242647,   1.00000000),rgb = (  0.64313725,   0.97769608,   1.00000000),rgb = (  0.63921569,   0.98308824,   1.00000000),rgb = (  0.63529412,   0.98860294,   1.00000000),rgb = (  0.63137255,   0.99424020,   1.00000000),rgb = (  0.62745098,   1.00000000,   1.00000000),rgb = (  0.62941176,   1.00000000,   0.99411765),rgb = (  0.63149510,   1.00000000,   0.98811275),rgb = (  0.63370098,   1.00000000,   0.98198529),rgb = (  0.63602941,   1.00000000,   0.97573529),rgb = (  0.63848039,   1.00000000,   0.96936275),rgb = (  0.64105392,   1.00000000,   0.96286765),rgb = (  0.64375000,   1.00000000,   0.95625000),rgb = (  0.64656863,   1.00000000,   0.94950980),rgb = (  0.64950980,   1.00000000,   0.94264706),rgb = (  0.65257353,   1.00000000,   0.93566176),rgb = (  0.65575980,   1.00000000,   0.92855392),rgb = (  0.65906863,   1.00000000,   0.92132353),rgb = (  0.66250000,   1.00000000,   0.91397059),rgb = (  0.66605392,   1.00000000,   0.90649510),rgb = (  0.66973039,   1.00000000,   0.89889706),rgb = (  0.67352941,   1.00000000,   0.89117647),rgb = (  0.67745098,   1.00000000,   0.88333333),rgb = (  0.68149510,   1.00000000,   0.87536765),rgb = (  0.68566176,   1.00000000,   0.86727941),rgb = (  0.68995098,   1.00000000,   0.85906863),rgb = (  0.69436275,   1.00000000,   0.85073529),rgb = (  0.69889706,   1.00000000,   0.84227941),rgb = (  0.70355392,   1.00000000,   0.83370098),rgb = (  0.70833333,   1.00000000,   0.82500000),rgb = (  0.71323529,   1.00000000,   0.81617647),rgb = (  0.71825980,   1.00000000,   0.80723039),rgb = (  0.72340686,   1.00000000,   0.79816176),rgb = (  0.72867647,   1.00000000,   0.78897059),rgb = (  0.73406863,   1.00000000,   0.77965686),rgb = (  0.73958333,   1.00000000,   0.77022059),rgb = (  0.74522059,   1.00000000,   0.76066176),rgb = (  0.75098039,   1.00000000,   0.75098039),rgb = (  0.75686275,   1.00000000,   0.74117647),rgb = (  0.76286765,   1.00000000,   0.73125000),rgb = (  0.76899510,   1.00000000,   0.72120098),rgb = (  0.77524510,   1.00000000,   0.71102941),rgb = (  0.78161765,   1.00000000,   0.70073529),rgb = (  0.78811275,   1.00000000,   0.69031863),rgb = (  0.79473039,   1.00000000,   0.67977941),rgb = (  0.80147059,   1.00000000,   0.66911765),rgb = (  0.80833333,   1.00000000,   0.65833333),rgb = (  0.81531863,   1.00000000,   0.64742647),rgb = (  0.82242647,   1.00000000,   0.63639706),rgb = (  0.82965686,   1.00000000,   0.62524510),rgb = (  0.83700980,   1.00000000,   0.61397059),rgb = (  0.84448529,   1.00000000,   0.60257353),rgb = (  0.85208333,   1.00000000,   0.59105392),rgb = (  0.85980392,   1.00000000,   0.57941176),rgb = (  0.86764706,   1.00000000,   0.56764706),rgb = (  0.87561275,   1.00000000,   0.55575980),rgb = (  0.88370098,   1.00000000,   0.54375000),rgb = (  0.89191176,   1.00000000,   0.53161765),rgb = (  0.90024510,   1.00000000,   0.51936275),rgb = (  0.90870098,   1.00000000,   0.50698529),rgb = (  0.91727941,   1.00000000,   0.49448529),rgb = (  0.92598039,   1.00000000,   0.48186275),rgb = (  0.93480392,   1.00000000,   0.46911765),rgb = (  0.94375000,   1.00000000,   0.45625000),rgb = (  0.95281863,   1.00000000,   0.44325980),rgb = (  0.96200980,   1.00000000,   0.43014706),rgb = (  0.97132353,   1.00000000,   0.41691176),rgb = (  0.98075980,   1.00000000,   0.40355392),rgb = (  0.99031863,   1.00000000,   0.39007353),rgb = (  1.00000000,   1.00000000,   0.37647059),rgb = (  1.00000000,   0.99019608,   0.37254902),rgb = (  1.00000000,   0.98026961,   0.36862745),rgb = (  1.00000000,   0.97022059,   0.36470588),rgb = (  1.00000000,   0.96004902,   0.36078431),rgb = (  1.00000000,   0.94975490,   0.35686275),rgb = (  1.00000000,   0.93933824,   0.35294118),rgb = (  1.00000000,   0.92879902,   0.34901961),rgb = (  1.00000000,   0.91813725,   0.34509804),rgb = (  1.00000000,   0.90735294,   0.34117647),rgb = (  1.00000000,   0.89644608,   0.33725490),rgb = (  1.00000000,   0.88541667,   0.33333333),rgb = (  1.00000000,   0.87426471,   0.32941176),rgb = (  1.00000000,   0.86299020,   0.32549020),rgb = (  1.00000000,   0.85159314,   0.32156863),rgb = (  1.00000000,   0.84007353,   0.31764706),rgb = (  1.00000000,   0.82843137,   0.31372549),rgb = (  1.00000000,   0.81666667,   0.30980392),rgb = (  1.00000000,   0.80477941,   0.30588235),rgb = (  1.00000000,   0.79276961,   0.30196078),rgb = (  1.00000000,   0.78063725,   0.29803922),rgb = (  1.00000000,   0.76838235,   0.29411765),rgb = (  1.00000000,   0.75600490,   0.29019608),rgb = (  1.00000000,   0.74350490,   0.28627451),rgb = (  1.00000000,   0.73088235,   0.28235294),rgb = (  1.00000000,   0.71813725,   0.27843137),rgb = (  1.00000000,   0.70526961,   0.27450980),rgb = (  1.00000000,   0.69227941,   0.27058824),rgb = (  1.00000000,   0.67916667,   0.26666667),rgb = (  1.00000000,   0.66593137,   0.26274510),rgb = (  1.00000000,   0.65257353,   0.25882353),rgb = (  1.00000000,   0.63909314,   0.25490196),rgb = (  1.00000000,   0.62549020,   0.25098039),rgb = (  1.00000000,   0.61176471,   0.24705882),rgb = (  1.00000000,   0.59791667,   0.24313725),rgb = (  1.00000000,   0.58394608,   0.23921569),rgb = (  1.00000000,   0.56985294,   0.23529412),rgb = (  1.00000000,   0.55563725,   0.23137255),rgb = (  1.00000000,   0.54129902,   0.22745098),rgb = (  1.00000000,   0.52683824,   0.22352941),rgb = (  1.00000000,   0.51225490,   0.21960784),rgb = (  1.00000000,   0.49754902,   0.21568627),rgb = (  1.00000000,   0.48272059,   0.21176471),rgb = (  1.00000000,   0.46776961,   0.20784314),rgb = (  1.00000000,   0.45269608,   0.20392157),rgb = (  1.00000000,   0.43750000,   0.20000000),rgb = (  1.00000000,   0.42218137,   0.19607843),rgb = (  1.00000000,   0.40674020,   0.19215686),rgb = (  1.00000000,   0.39117647,   0.18823529),rgb = (  1.00000000,   0.37549020,   0.18431373),rgb = (  1.00000000,   0.35968137,   0.18039216),rgb = (  1.00000000,   0.34375000,   0.17647059),rgb = (  1.00000000,   0.32769608,   0.17254902),rgb = (  1.00000000,   0.31151961,   0.16862745),rgb = (  1.00000000,   0.29522059,   0.16470588),rgb = (  1.00000000,   0.27879902,   0.16078431),rgb = (  1.00000000,   0.26225490,   0.15686275),rgb = (  1.00000000,   0.24558824,   0.15294118),rgb = (  1.00000000,   0.22879902,   0.14901961),rgb = (  1.00000000,   0.21188725,   0.14509804),rgb = (  1.00000000,   0.19485294,   0.14117647),rgb = (  1.00000000,   0.17769608,   0.13725490),rgb = (  1.00000000,   0.16041667,   0.13333333),rgb = (  1.00000000,   0.14301471,   0.12941176),rgb = (  1.00000000,   0.12549020,   0.12549020),rgb = (  0.98627451,   0.12156863,   0.12156863),rgb = (  0.97242647,   0.11764706,   0.11764706),rgb = (  0.95845588,   0.11372549,   0.11372549),rgb = (  0.94436275,   0.10980392,   0.10980392),rgb = (  0.93014706,   0.10588235,   0.10588235),rgb = (  0.91580882,   0.10196078,   0.10196078),rgb = (  0.90134804,   0.09803922,   0.09803922),rgb = (  0.88676471,   0.09411765,   0.09411765),rgb = (  0.87205882,   0.09019608,   0.09019608),rgb = (  0.85723039,   0.08627451,   0.08627451),rgb = (  0.84227941,   0.08235294,   0.08235294),rgb = (  0.82720588,   0.07843137,   0.07843137),rgb = (  0.81200980,   0.07450980,   0.07450980),rgb = (  0.79669118,   0.07058824,   0.07058824),rgb = (  0.78125000,   0.06666667,   0.06666667),rgb = (  0.76568627,   0.06274510,   0.06274510),rgb = (  0.75000000,   0.05882353,   0.05882353),rgb = (  0.73419118,   0.05490196,   0.05490196),rgb = (  0.71825980,   0.05098039,   0.05098039),rgb = (  0.70220588,   0.04705882,   0.04705882),rgb = (  0.68602941,   0.04313725,   0.04313725),rgb = (  0.66973039,   0.03921569,   0.03921569),rgb = (  0.65330882,   0.03529412,   0.03529412),rgb = (  0.63676471,   0.03137255,   0.03137255),rgb = (  0.62009804,   0.02745098,   0.02745098),rgb = (  0.60330882,   0.02352941,   0.02352941),rgb = (  0.58639706,   0.01960784,   0.01960784),rgb = (  0.56936275,   0.01568627,   0.01568627),rgb = (  0.55220588,   0.01176471,   0.01176471),rgb = (  0.53492647,   0.00784314,   0.00784314),rgb = (  0.51752451,   0.00392157,   0.00392157),rgb = (  0.50000000,   0.00000000,   0.00000000)}
}

\addto\captionsenglish{}
\definecolor{mittelblau}{RGB}{0, 126, 198}
\definecolor{violettblau}{cmyk}{0.9, 0.6, 0, 0}
\definecolor{rot}{RGB}{238, 28 35}
\definecolor{apfelgruen}{RGB}{140, 198, 62}
\definecolor{gelb}{RGB}{1, 221, 0}
\definecolor{orange}{RGB}{244, 111, 33}
\definecolor{pink}{RGB}{237, 0, 140}
\definecolor{lila}{RGB}{128, 10, 145}
\definecolor{hellgrau}{RGB}{224, 224, 224}
\definecolor{mittelgrau}{RGB}{128, 128, 128}
\definecolor{dunkelgrau}{RGB}{80,80,80}
\definecolor{anthrazit}{RGB}{19, 31, 31}

\definecolor{myblue}{RGB}{80,80,160} 
\definecolor{mygreen}{RGB}{80,160,80}
\definecolor{myorgange}{RGB}{204,102,0}
\definecolor{lightblue}{RGB}{51,153,255}

\makeatother

\usepackage{babel}
\providecommand{\theoremname}{Theorem}

\begin{document}

\title{Achievable Rate Region for Iterative Multi-User Detection via Low-cost Gaussian Approximation}

\author{Xiaojie~Wang,~\IEEEmembership{Student~Member,~IEEE,}~Chulong~Liang,~Li~Ping,~\IEEEmembership{Fellow,~IEEE,}
and~Stephan~ten~Brink,~\IEEEmembership{Senior~Member,~IEEE}
\thanks{Part of the results is to appear in the proceedings of the IEEE International Symposium on Information Theory 2019, Paris, France \cite{XJWISIT19}.}
\thanks{X. J. Wang and S. ten Brink are with Institute of Telecommunications, Pfaffenwaldring 47, University of Stuttgart, 70569 Stuttgart, Germany (e-mail: \{wang, tenbrink\}@inue.uni-stuttgart.de).}
\thanks{C. Liang and L. Ping are with the Department of Electronic Engineering, City University of Hong Kong, Hong Kong SAR, China (e-mail: \{chuliang, eeliping\}@cityu.edu.hk).}}
\maketitle
\begin{abstract}
We establish a multiuser extrinsic information transfer (EXIT) chart
area theorem for the interleave-division multiple access (IDMA) scheme,
a special form of superposition coding, in multiple access channels
(MACs). A low-cost multi-user detection (MUD) based on the Gaussian approximation
(GA) is assumed. The evolution of mean-square errors (MSE) of the
GA-based MUD during iterative processing is studied. We show that the
$K$-dimensional tuples formed by the MSEs of $K$ users constitute
a conservative vector field. The achievable rate is a potential function
of this conservative field, so it is the integral along any path in
the field with value of the integral solely determined by the two path terminals.
Optimized codes can be found given the integration paths in the MSE fields
by matching EXIT type functions.
The above findings imply that i) low-cost GA detection can provide
near capacity performance, ii) the sum-rate capacity can be achieved
independently of the integration path in the MSE fields; and iii) the
integration path can be an extra degree of freedom for code design.
\end{abstract}


\section{Introduction}\label{sec:intro}

Consider a multiple access channel (MAC) with $K$ users. 
The MAC capacity region is bounded by $2^{K}-1$ constraints and determined by a tuple of user rates $R_{k},$ $1\leq k\leq K$ \cite{CoverITBook06,GamalNetInfTh}.
To achieve arbitrary points of the capacity region, joint detection
and decoding is required, which has prohibitively high complexity exponential to $K$. 
Theoretically, successive interference cancellation
(SIC) together with time-sharing or rate-splitting can achieve the
entire capacity region \cite{DTseBookFund}. SIC involves subtraction
of successfully detected signals. If practical forward error control
(FEC) codes are used, each subtraction incurs an overhead in
terms of either power or rate loss relative to an ideal capacity achieving
code \cite[Fig. 13.3]{YHuLPNOMAbook}. Such overheads accumulate during
SIC steps, moving its performance away from the capacity.
Also, both time-sharing and rate-splitting
involve segmenting a data frame of a user into several sub-frames.
In practice, the length of a coding frame is restricted the latency requirement. 
Frame segmentation results in shorter sub-frames and so reduced coding gain for a practical turbo or low-density parity-check (LDPC) type code \cite{TomMCTBook}, which further worsens the losses of accumulation.

Iterative detection \cite{SchlegelCDMA,HanzoCEMUD,PHoeherIC,ReedTurbo} can alleviate the loss accumulation problem using
soft cancellations instead of hard subtraction. A turbo or LDPC code involving
iterative detection can be optimized by matching the so-called extrinsic
information transfer (EXIT) functions of two local processors \cite{tBAllerton01,RichardsonLDPCDesignTIT01}.
In a single-user point-to-point channel, such matching can offer near
capacity performance, as shown by the area properties \cite{AKtBAreaThBEC04IT,BNMSEChart07IT}.

Interleave-division multiple-access (IDMA) is a low-cost transmission
scheme for MACs \cite{LipingIDMATWC06}. A Gaussian approximation (GA)
of cross-user interference is key to a low-cost IDMA detection technique.
The per-user complexity of a GA-based MUD remains roughly the same for all K. For comparison, the complexity of a
standard \textit{a posteriori} probability (APP) based multi-user detector~(MUD) is exponential in $K$ \cite{LipingIDMATWC06}.

A question naturally arises: At such low cost, what is the achievable
performance of IDMA under GA-based MUD? Some partial answers to this question
are available. It is shown that IDMA is capacity approaching when
all users see the same channel \cite{ChulongCLIDMASCMA}. However,
for the general MAC system, the code design for IDMA becomes difficult when different users see different channels. In the latter case, to achieve the entire capacity region, different coding rates or different power levels are generally required. Previous works on IDMA focus on the achievability of some special points in the MAC capacity region  \cite{KLiEXITTurboMUTWC05,GSongISIT15,LLiuTSP18,XJWISTC18,XJWangSubmitedTCOM} and/or other aspects, e.g., power control etc.\cite{KLiIDMAOptTWC07,KusumeIDMA12,SongISTC18,CEiDMA,RateOptIDMA}. 
To the best of our knowledge, no previous work has shown that IDMA under the GA-based MUD can achieve the entire MAC capacity region.

This paper provides a comprehensive analysis of the achievable performance
of IDMA under GA-based MUD. We approach the problem based on multi-dimensional
curve matching. Let $v_{k}$ be the mean-square error~(MSE) (i.e., the
variance) for the GA-based MUD for user $k$, with $v_{k}=0$ indicating
perfect decoding. Using the relationship between mutual information (MI) and minimum MSE~(MMSE) derived in \cite{GuoSVMMSEmIITI05,BNMSEChart07IT},
we show that the achievable sum-rate can be evaluated using a line
integral along a valid path in the $K$-dimensional vector field $\boldsymbol{v}=\left[v_{1},v_{2},\cdots,v_{K}\right]^{T}$.
A main finding of this paper is that the integral is path-independent and its value is solely determined by the two terminations.
The path independence property greatly simplifies the code optimization problem.

The contributions of this paper are summarized as follows.
\begin{itemize}
\item A low-cost GA-based MUD can provide near optimal performance.
In particular, it is provably capacity-achieving for Gaussian signaling.
\item Relative to Gaussian signaling, the loss due to finite modulation can be made arbitrarily small using a superposition coded modulation (SCM) technique.
\item FEC codes optimized for single-user channels may not be good choices
for MACs. 
The FEC codes should be carefully designed to match MUD, which facilitates iterative detection. We
will provide examples for the related code design.
\item A multi-user area theorem of EXIT chart is established for the code design. 
We show that the sum-rate capacity is a potential function in the MSE field formed
by $\boldsymbol{v}$, which leads to the path independence property.
All points of the MAC capacity region are achievable using only
one FEC code per user. This avoids the loss related to the frame segmentation
of SIC as aforementioned.
\item The above results can be extended to MIMO MAC channels straightforwardly. We will provide simulation results to show that properly designed
IDMA can approach the sum-rate MAC capacity for different decoding paths in the MSE field within $\unit[1]{dB}$ .
\end{itemize}

This paper is structured as follows. In Sec.~\ref{sec:IDMA}, we present the multiuser
iterative detection and decoding scheme in IDMA along with the matching
condition. Then, we derive the achievable rates of IDMA and show its
implication in code design in single antenna setup in Sec.~\ref{sec:Rate}. The
achievable rate analysis is further extend to MIMO cases in Sec.~\ref{sec:MIMO}.
Sec.~\ref{sec:Results} provides code design examples and numerical results verifying
our theorems. Finally, Sec.~\ref{sec:conclusion} concludes the paper.

\section{Iterative IDMA Receiver}\label{sec:IDMA}

Consider a general $K$-user MAC system, which is described by
\begin{equation}
y={\displaystyle \sum_{k=1}^{K}}\sqrt{P_{k}}h_{k}x_{k}+n \tag{1}\label{eq:SysModel}
\end{equation}
where $P_{k}$ denotes the received signal strength of the $k$th user's
signal, $h_{k}$ denotes the fading coefficients of the user, $x_{k}$
is the $k$th transmit signal and $n$ is the additive (circularly symmetric complex) white Gaussian noise~(AWGN) with zero mean and unit variance, i.e., $\mathcal{CN}\left(0,\sigma^{2}=1\right)$. We consider complex-valued signals throughout the paper if not otherwise stated.
Practically, a sequence of symbols $y$, forming one or multiple codewords $\mathbf{y}$, is received.  

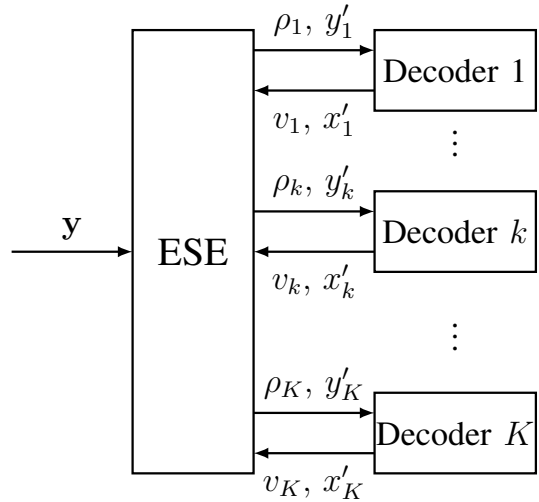
\begin{figure}[tbh]
\begin{center}
\resizebox {.45\linewidth} {!} {

\begin{tikzpicture} [>=latex]
\draw [thick] (1,3) rectangle (3,2);
\node at (2,2.5) {Decoder 1};
\draw [thick] (1,1) rectangle (3,0);
\node at (2,0.5) {Decoder $k$};
\node at (2,-0.7) {$\vdots$};
\node at (2, 1.7) {$\vdots$};
\draw [thick] (1,-1.5) rectangle (3,-2.5);
\node at (2,-2) {Decoder $K$};
\draw [thick] (-2,3) rectangle (-0.5,-2.5);
\node at (-1.25,0.25) {\large ESE};
\draw [thick,->](-3.5,0.25) -- (-2,0.25) node [midway, above] {$\mathbf{y}$};
\draw [thick,->](-0.5,2.75) -- (1,2.75) node [midway, above] {$\rho_{1},\,y'_{1}$};
\draw [thick,<-](-0.5,2.25) -- (1,2.25) node [midway, below] {$v_{1},\,x'_{1}$};
\draw [thick,->](-0.5,0.75) -- (1,0.75) node [midway, above] {$\rho_{k},\,y'_{k}$};
\draw [thick,<-](-0.5,0.25) -- (1,0.25) node [midway, below] {$v_{k},\,x'_{k}$};
\draw [thick,->](-0.5,-1.75) -- (1,-1.75) node [midway, above] {$\rho_{K},\,y'_{K}$};
\draw [thick,<-](-0.5,-2.25) -- (1,-2.25) node [midway, below] {$v_{K},\,x'_{K}$};
\end{tikzpicture}

}
\end{center}
\caption{The iterative multi-user detection and decoding model in IDMA.\label{fig:The-decoder-model-MU}}
\end{figure}

The iterative receiver is depicted in Fig. \ref{fig:The-decoder-model-MU}.
The elementary signal estimator (ESE) module has access to the channel
observation $y$ and feedbacks $x'_{k}$ from all the users' decoders.
It performs the so called soft interference cancellation (SoIC) and output signals with reduced interference. 
Each decoder (DEC) performs decoding for a particular user while treating the residual signals of other users as noise. 
Through the iterative message passing between the ESE and DECs, the wanted signals are refined and interference suppressed progressively. 
More details on the ESE DECs are given below.  
For convenience of discussions, we will assume that $x_i$ are modulated using binary phase shift keying (BPSK). 


\subsection{ESE functions}\label{subsec:IDMAESE}
The function of the ESE (for elementary signal estimation) module in Fig.~\ref{fig:The-decoder-model-MU} is interference cancellation
The outputs of ESE are a sequence of $y_{k}$ with 
\begin{align*}
y_{k} &= y - \tilde{z}_{k}=\sqrt{P_{k}}h_{k}x_{k}+ z_{k} \tag{2a}\label{eq:GaussAssum1} \\ 
z_{k} &= {\displaystyle \sum_{i=1, i\ne k}^{K}}\sqrt{P_{i}}h_{i}\left(x_{i} - \hat{x}_{i}\right)+n.\tag{2b}\label{eq:MAI}
\end{align*}
Here, $y_{k}$ is obtained from $y$ in \eqref{eq:SysModel} by canceling out the mean of the interference $\tilde{z}_{k}={\displaystyle \sum_{i=1, i\ne k}^{K}}\sqrt{P_{i}}h_{i}\hat{x}_{i}$ based on the feedback of the channel decoders.
The soft symbol estimates $\hat{x}_{i}$ are generated by feedbacks from the users' channel decoders. 
For instance with BPSK signaling, $\hat{x}_{i}=\mathrm{tanh}\left(\frac{L_{i}}{2}\right)$ with $L_{i}$ being the  log-likelihood ratio (LLR) after decoding. For higher modulation schemes, the soft symbol estimates can be obtained by \cite[eqn. 3a]{XYuanITI14}.
The feedback from channel decoders will be discussed later in Sec.~\ref{subsec:IDMADEC}. 
The term $z_k$ in \eqref{eq:MAI} is comprised of AWGN and residual multi-user interference. 
To reduce complexity, we will adopt a Gaussian approximation (GA) assuming that $z_k$ is Gaussian-distributed with zero mean and variance $\sigma_{z,k}^{2}$, i.e., $\mathcal{CN} (0, \sigma_{z,k}^{2})$
\footnote{The Gaussian assumption is valid for a large number of users with arbitrary independently transmitted symbols $x_{i}$ as the consequence of the central limit theorem or if the transmit signals $x_{i}$ are Gaussian by themselves.}. 
From \eqref{eq:MAI}, we obtain 
\begin{align*}
\sigma_{z,k}^{2} &= {\displaystyle \sum_{i=1, i\ne k}^{K}} P_{i}\left|h_{i}\right|^{2}v_{i}+ \sigma^{2} \tag{3a}\label{eq:GaussAssum1} \\ 
\end{align*}
where the MSE of the symbol estimates $v_{i} = \mathrm{E}\left[ \left| x_{i} - \hat{x}_{i}\right|^{2} \right]$ is to characterize the quality of the decoder feedback $\hat{x}_{i}$. 
The complexity in \eqref{eq:GaussAssum1} can be reduced by a sum-and-minus trick by noting that $\tilde{z}_{k} = \Sigma - \sqrt{P_{k}}h_{k}\hat{x}_{k}$ where $\Sigma ={\displaystyle \sum_{i=1}^{K}}\sqrt{P_{i}}h_{i}\hat{x}_{i}$. Here $\Sigma$ is common to all users, so its cost can be shared. The per user cost for \eqref{eq:GaussAssum1} thus does not grow with K.
The quality of the ESE output for user $k$ can be measured by the signal-to-noise ratio (SNR) offered by $y_{k}$ in \eqref{eq:GaussAssum1}
\begin{equation}
\rho_{k}=\frac{P_{k}\left|h_{k}\right|^{2}}{\sigma_{z,k}^{2}}=\frac{P_{k}\left|h_{k}\right|^{2}}{{\displaystyle \sum_{i=1,i\ne k}^{K}}P_{i}\left|h_{i}\right|^{2}v_{i}+\sigma^{2}},\:\forall k=1,2,\ldots,K.\tag{3b}\label{eq:SINR}
\end{equation}
Assume that the average power of $x_i$ is normalized to 1. Then $v_{i}=1$ in the first iteration, meaning no a prior information about $x_i$. During the iterative detection, $v_i$ will be updated using decoder output (See the discussion in Sec. II-B below).
For $K$ users, we express \eqref{eq:SINR} in a vector form as
\begin{equation}
\boldsymbol{\rho}=\phi\left(\boldsymbol{v}\right)\tag{3c}\label{eq:SINRvec}
\end{equation}
where $\boldsymbol{\rho}=\left[\rho_{1},\rho_{2},\cdots,\rho_{K}\right]^{T}$
and $\boldsymbol{v}=\left[v_{1},v_{2},\cdots,v_{K}\right]^{T}$. Due
to the fact that the MSE is bounded by $0\leq v_{i}\leq\mathrm{E}\left[\left|x_{i}\right|^{2}\right]=1,$we
obtain that the SNR is also bounded by
\begin{equation}
\rho_{k,\mathrm{min}}=\frac{P_{k}\left|h_{k}\right|^{2}}{{\displaystyle \sum_{i=1,i\ne k}^{K}}P_{i}\left|h_{i}\right|^{2}+\sigma^{2}}\leq\rho_{k}\leq\frac{P_{k}\left|h_{k}\right|^{2}}{\sigma^{2}}=\rho_{k,\mathrm{max}}. \tag{3d}\label{eq:SINRrange}
\end{equation}

We will view (3) as a transfer function from $\boldsymbol{v}$ to $\boldsymbol{\rho}$. 


\subsection{DEC functions}\label{subsec:IDMADEC}
The refined signals $y_{k}$ in (2) generated by the ESE are forwarded to the DECs. The latter consists of $K$ local decoders (DECs, see Fig.~\ref{fig:The-decoder-model-MU}) performing extrinsic decoding based on $y_{k}$ with SNR ${\rho}_{k}$.
To reduce complexity, we will adopt a Gaussian approximation (GA) that $z_{k}$ is Gaussian-distributed with zero mean and variance $\sigma_{z,k}^{2}$. Then the standard decoding operations \cite{RichardsonLDPCDesignTIT01, TomMCTBook} can be applied to the local decoders.
The outputs of an APP decoder are extrinsic messages that are assumed to resemble observations from the AWGN channel, i.e., 
\begin{equation}
x'_{k}=x_{k}+w_{k}\tag{4}\label{eq:GaussAssum2}
\end{equation}
where $w_{k}$ follows a Gaussian-distribution $\mathcal{CN}\left(0,\sigma_{w,k}^{2}\right)$.
Let the MSEs for $\hat{x}_{k}$ be ${v}_k$ after decoding. 
The MSEs ${v}_k$ are also the MMSE of the conditional mean estimator due to APP decoding, thus, we define a transfer function for DEC $k$  as
\begin{equation}
{v}_{k}=\mathrm{E}\left[\left|x_{k}-\mathrm{E}\left[\left.x_{k}\right|x'_{k}\right]\right|^{2}\right]=\psi_{k}\left({\rho}_{k}\right),0\leq {v}_{k}\leq1,\:\forall k=1,2,\ldots,K.\tag{5a}\label{eq:DecFuncDef}
\end{equation}
Or in a vector form for the overall DEC
\begin{equation}
{\boldsymbol{v}}=\psi\left({\boldsymbol{\rho}}\right).\tag{5b}\label{eq:DecFuncVec}
\end{equation}
In general, unlike $\phi\left(\cdot\right)$ in \eqref{eq:SINRvec}, we do not have an explicit expression for $\psi\left(\cdot\right)$ in \eqref{eq:DecFuncVec}, but it can be numerically measured. The details can be found in \cite{tenBrinkEXITTC01}.

In general, $\psi_{k}\left({\rho}_{k}\right)$ can be 
\begin{align*}
v_{k} =
\begin{cases}
1, &\quad \rho\le\rho_{k,\mathrm{min}} \tag{5c} \\ 
\psi_{k}\left(\rho_{k}\right),&\quad \rho_{k,\mathrm{max}}\le\rho\le\rho_{k,\mathrm{min}}\\
0 &\quad \rho\ge\rho_{k,\mathrm{max}}\\
\end{cases}
\end{align*} 
Here the first case of $v_{k}=1$ is for the boundary condition $\rho\le\rho_{k,\mathrm{min}}$ in \eqref{eq:SINRrange} at the start of the iterative detection. On the other hand, the last case of $v_{k}=0$ is for the boundary condition  $\rho\ge\rho_{k,\mathrm{max}}$ in \eqref{eq:SINRrange} at the end of the iterative detection when all interference has been perfectly canceled  out and perfect decoding is assumed to be achievable at this point.

To track the convergence behavior of the iterations between ESE and DECs, we write the SNR ${\boldsymbol{\rho}}$ and MSE ${\boldsymbol{v}}$ vector as functions of an iteration variable $t$  as 
\begin{align}
\boldsymbol{v} =\boldsymbol{v} \left(t\right) \quad \mathrm{and } \, \boldsymbol{\rho} =\boldsymbol{\rho} \left(t\right) \tag{5d}\label{eq:IterationIndex}
\end{align}
Let $t_{0}$ and $t_{\infty}$ denote the start and end of the iterative processing, we require that 
\begin{align}
{\boldsymbol{v}} \left(t_{0}\right)& =\psi\left({\boldsymbol{\rho}}\left(t_{0}\right)\right)=\boldsymbol{1}\tag{5e}\label{eq:DecIni}\\
{\boldsymbol{v}} \left(t_{\infty}\right) & =\psi\left({\boldsymbol{\rho}}\left(t_{\infty}\right)\right)=\boldsymbol{0}\tag{5f}\label{eq:DecConv}
\end{align}
since we are interested in the error-free decoding cases.

\subsection{Matching condition}\label{subsec:IDMAMatch}

We will say that the ESE and DEC functions are matched if the following condition is met
\begin{equation}
\psi\left(\boldsymbol{\rho}\left(t\right)\right)=\phi^{-1}\left(\boldsymbol{\rho}\left(t\right)\right).\tag{6}\label{eq:MatchCondLine}
\end{equation}
Note that the matching condition in \eqref{eq:MatchCondLine} is along a $K$-dimensional line given by $\boldsymbol{\rho}\left(t\right)$. 
It is not required to match $\psi\left(\boldsymbol{\rho}\right)$ and $\phi\left(\boldsymbol{\rho}\right)$ in the entire $K$ dimensional space, i.e., requiring $\psi\left(\boldsymbol{\rho}\right)=\phi^{-1}\left(\boldsymbol{\rho}\right)$. 
The line matching in \eqref{eq:MatchCondLine} is  much easier. 
We will show that such line matching achieves the MAC capacity (see  Sec.~\ref{subsec:RateGauss}).


\section{Achievable rates}\label{sec:Rate}

The fundamental relation between achievable rate and MMSE in AWGN channels $y=x+n$ is found by Guo et. al. \cite{GuoSVMMSEmIITI05} as 

\begin{align*}
R\left(\mathrm{snr}\right)={\displaystyle \int_{0}^{\mathrm{snr}}}\mathrm{mmse}\left(\rho\right)d\rho.
\end{align*}
for any input distribution of $x$. 
The above result is  extended to iterative decoding in \cite{BNMSEChart07IT}.
Following \cite{BNMSEChart07IT} and also \cite{GuoSVMMSEmIITI05, XYuanITI14}, the achievable rate for user $k$ using GA-based MUD is given by
\begin{equation}
R_{k}={\displaystyle \int_{0}^{\infty}}f\left(\rho_{k}+f^{-1}\left(v_{k}\right)\right)d\rho_{k},\:\forall k=1,2,\ldots,K.\tag{7}\label{eq:UserRateIntGeneral}
\end{equation}
where $f\left(\rho\right)=v$ is the achievable MMSE for a given constellation of $x$ by observing $y$ at the SNR of $\rho$. 
Intuitively, $\rho_{k}$ and $f^{-1}\left(v_{k}\right)$ give, respectively, the SNRs related to the input and extrinsic messages of the $k$th DEC. Hence $\rho_{k}+f^{-1}\left(v_{k}\right)$ represents the overall SNR after combining these two messages. 

\subsection{Gaussian alphabets}\label{subsec:RateGauss}

We first consider the case when $x_i$ are Gaussian distributed.
This can be approximated by using e.g., superposition coded modulation
(SCM) \cite{LPSCMJSAC09,JYuanSCM}. The MMSE for Gaussian signals is given by $f\left(\rho\right)=\frac{1}{1+\rho},$
and with \eqref{eq:UserRateIntGeneral} the achievable rates can be expressed as  
\begin{equation*}
R_{k}={\displaystyle \int_{0}^{\infty}}\frac{1}{\rho_{k}\left(t\right)+v_{k}^{-1}\left(t\right)}d\rho_{k}\left(t\right), \,\forall k=1,2,\ldots,K. \tag{7a}\label{eq:Rate7a}
\end{equation*}
Here $v_{k}\left(t\right)$ and $\rho_{k}\left(t\right)$ are related by the function in \eqref{eq:SINR} and (5), and they are expressed as the functions of $t$, as introduced in \eqref{eq:IterationIndex}. In Appendix A, we will consider \eqref{eq:SINR} and (5) and rewrite \eqref{eq:Rate7a} into the following form: 
\begin{equation}
R_{k} = -{\displaystyle \int}_{v_{k}=1}^{v_{k}=0}\frac{g_{k}}{ {\displaystyle \sum_{i=1}^{K}g_{i}v_{i}\left(t\right)} +\sigma^{2}}dv_{k}\left(t\right) =-{\displaystyle \int}_{v_{k}=1}^{v_{k}=0}\frac{g_{k}}{\mathbf{g}^{T}\boldsymbol{v}\left(t\right)+\sigma^{2}}dv_{k}\left(t\right) 
\forall k=1,2,\ldots,K
\tag{7b}\label{eq:UserRateGaussian}
\end{equation}
where $\mathbf{g}^{T}=\left[P_{1}\left|h_{1}\right|^{2},P_{2}\left|h_{2}\right|^{2},\cdots,P_{K}\left|h_{K}\right|^{2}\right]^{T}$
contains the powers of all users, $\boldsymbol{v}=\left[v_{1},v_{2},\cdots,v_{K}\right]^{T}$
and $g_{k}=P_{k}\left|h_{k}\right|^{2}$ denotes the $k$th element
of vector $\mathbf{g}$.
Note that the achievable rate expression in \eqref{eq:UserRateGaussian} depends on multiple variables $v_1$, $v_2$, $\cdots$ and $v_{K}$, i.e., the evolution of the MSEs of all the DECs. This can be intuitively explained by the iterative soft interference cancellation of the ESE based on other users' DEC feedbacks. 


Hence, the achievable sum-rate of all users  can be written as
\begin{align}
R_{\mathrm{sum}} & ={\displaystyle \sum_{k=1}^{K}}R_{k}=-{\displaystyle \int}_{\boldsymbol{v}\left(t\right)}\frac{\mathbf{g}}{\mathbf{g}^{T}\boldsymbol{v}\left(t\right)+\sigma^{2}}\cdot d\boldsymbol{v}\left(t\right)\tag{8a}\label{eq:LineIntRsum-1}
\end{align}
where \eqref{eq:LineIntRsum-1} denotes a line integral defined by
$\boldsymbol{v}\left(t\right),\:t\in\left[t_{0},t_{\infty}\right]$.
Notice that the line $\boldsymbol{v}\left(t\right)$ is determined by the evolution of the MSE vector $\boldsymbol{v}$ of all DECs. 
We recall that the terminals of the line are given in \eqref{eq:DecIni} and \eqref{eq:DecConv} as $\boldsymbol{v}\left(t_{0}\right)=\mathbf{1}$ and  $\boldsymbol{v}\left(t_{\infty}\right)=\mathbf{0}$.
It can be verified that the integrands constitute a gradient of
a scalar field (or potential function), i.e., $\frac{\mathbf{g}}{\mathbf{g}^{T}\boldsymbol{v}+\sigma^{2}} = \nabla_{\boldsymbol{v}}\mathrm{log}\left(\sigma^{2}+\mathbf{g}^{T}\boldsymbol{v}\right)$.
Thus, the achievable sum-rate can be written as
\begin{align*}
R_{\mathrm{sum}} & =-{\displaystyle \int_{L=\boldsymbol{v}\left(t\right)}}\left[\nabla\mathrm{log}\left(\sigma^{2}+\mathbf{g}^{T}\boldsymbol{v}\right)\right]\boldsymbol{v}'\left(t\right)dt\\
& = \mathrm{log}\left(\frac{\sigma^{2}+\mathbf{g}^{T}\boldsymbol{v}\left(t_{0}\right)}{\sigma^{2}+\mathbf{g}^{T}\boldsymbol{v}\left(t_{\infty}\right)}\right) \\
 & \overset{\eqref{eq:DecIni},\eqref{eq:DecConv}}{=}\mathrm{log}\left(1+\frac{\sum_{k=1}^{K}P_{k}\left|h_{k}\right|^{2}}{\sigma^{2}}\right)\tag{8b}\label{eq:SumrateCap}
\end{align*}
which is independent of the path taken for the integration. 
We note that the achievable rate in \eqref{eq:SumrateCap} coincides with the multi-user Shannon capacity.
In other words, any path with matched DEC functions can achieve the multi-user Shannon capacity. 
Therefore, the matching condition given in \eqref{eq:MatchCondLine} is a sufficient condition for achieving the sum-rate capacity.
It can be further verified that the matching condition also constitutes a necessary condition for achieving the multi-user Shannon capacity.
Consider the case $\boldsymbol{v}\left(t\right)<\phi^{-1}\left(\boldsymbol{\rho}\left(t\right)\right)$, 
then we have $R_{k}<-{\displaystyle \int}_{v_{k}=1}^{v_{k}=0}\frac{g_{k}}{\mathbf{g}^{T}\boldsymbol{v}+\sigma^{2}}dv_{k}$
and thus $R_{\mathrm{sum}}<\mathrm{log}\left(1+\frac{\sum_{k=1}^{K}P_{k}\left|h_{k}\right|^{2}}{\sigma^{2}}\right)$.
On the contrary, for the case $\psi\left(\boldsymbol{\rho}\left(t\right)\right)>\phi^{-1}\left(\boldsymbol{\rho}\left(t\right)\right)$,
error-free decoding is not possible.

This leads to the following theorem.
\begin{thm}
\label{thm:IDMA-is-capacity-achieving} The achievable sum-rate in
IDMA for any path $L\left(t\right):\boldsymbol{v}_{s}=\mathbf{1}\rightarrow\boldsymbol{v}_{e}=\mathbf{0}$
(starting from $\boldsymbol{v}_{s}=\mathbf{1}$ to $\boldsymbol{v}_{e}=\mathbf{0}$)
is given by the multiuser Shannon capacity
\begin{align*}
R_{\mathrm{sum}} & =-{\displaystyle \int}_{\boldsymbol{v}\left(t\right)}f\left(\boldsymbol{\rho}\left(t\right)+f^{-1}\left(\boldsymbol{v}\left(t\right)\right)\right)\cdot d\boldsymbol{\rho}\left(t\right)\\
 & =\mathrm{log}\left(1+\frac{\sum_{k=1}^{K}P_{k}\left|h_{k}\right|^{2}}{\sigma^{2}}\right)
\end{align*}
with the following assumptions
\begin{enumerate}
\item The exchanged messages during the iterative processing of the extrinsic and a priori
channel are observations from AWGN channels, given in \eqref{eq:GaussAssum1} and \eqref{eq:GaussAssum2}.
\item The channel decoder is APP (i.e, MAP) decoder.
\item The channel encoders and decoders use ``matched codes'' for a given path in the MMSE-field given in \eqref{eq:MatchCondLine}.
\end{enumerate}
\end{thm}

\textit{Remark:}
The assumptions used in Theorem 1 are common for turbo-type iterative receivers. It is generally accepted that these assumptions are sufficiently accurate for practical systems, based on which turbo and LDPC codes are designed in many modern communication systems \cite{tBAllerton01, tenBrinkLDPCEXITTC04}. Theorem 1 provides guidelines  for the design of FEC codes in the matching condition discussed in Sec.~\ref{subsec:IDMAMatch} for multi-user scenarios. Further, various  channel decoders such as BCJR and belief propagation (BP) are known for achieving APP performance.

\subsection{Finite alphabets}\label{subsec:RateFinite}

If the symbols $x_{i}\in\mathcal{S}_{i}$ are taken from finite alphabets
$\left|\mathcal{S}_{i}\right|<\infty$, the capacity formula, in general,
can not be expressed in closed-form. Notice that eq. \eqref{eq:UserRateIntGeneral}
is still valid, using the MMSE-formula for the underlying modulation
format $f\left(\rho\right)=v$. It is also well known that the loss
incurred by finite alphabets, compared with Gaussian, is negligible
in the low-SNR regime. Besides, the Gaussian capacity can be approached
by higher order modulations with shaping and/or SCM \cite{LPSCMJSAC09}.

We provide in the following an achievable rate analysis for IDMA with quadrature phase shift keying (QPSK) signaling.  
The achievable sum-rate can be written as
\begin{equation}
R_{\mathrm{sum}}={\displaystyle \int_{\boldsymbol{\rho}\left(t\right)}}f_Q\left(\boldsymbol{\rho}\left(t\right)+f_Q^{-1}\left(\boldsymbol{v}\left(t\right)\right)\right)d\boldsymbol{\rho}\left(t\right).\tag{9}\label{eq:UserRateIntQPSK}
\end{equation}
where $f_{Q}\left(\cdot\right)$ denotes the MMSE of QPSK, which is given by
\begin{align}\label{eq:MMSEQPSK}
f_{Q}(\rho) = 1 - \int_{-\infty}^{\infty} \tanh(\rho - y\sqrt{\rho})\frac{e^{-\frac{y^2}{2}}}{\sqrt{2\pi}} dy. \tag{10}
\end{align}
In AWGN channels, Gaussian signals are the hardest to estimate \cite{GuoMMSE}, i.e.,
\begin{align*}
f_{X}\left(\rho\right) \leq f_{G}\left(\rho\right)= \frac{1}{1+\rho}.
\end{align*}
for any input distribution $X$ with the same variance. 
Hence, the achievable rate with distributions other than Gaussian is essentially smaller.

\textit{Example}: The users are assumed to have same power $P$ and $h_{k}=1,\, \forall k$, modulation and coding scheme. For simplicity, we define $t_{0}=0$ and $t_{\infty}=1$ and consider the following integration path $\boldsymbol{\rho}\left(t\right)$
\begin{align*}
\boldsymbol{\rho}\left(t\right) = \frac{P}{\left(K-1\right)P+\sigma^{2}} \mathbf{1} + \left( \frac{P}{\sigma^{2}}-\frac{P}{\left(K-1\right)P+\sigma^{2}}\right)\cdot t \cdot\mathbf{1} \quad t \in [0,1]
\end{align*}
and with \eqref{eq:SINR} 
\begin{align*}
\boldsymbol{v}\left(t\right) =\frac{\left(1-t\right)\sigma^{2}}{\sigma^{2}+\left( K-1\right)P\cdot t}\cdot \mathbf{1}. \quad t \in [0,1]
\end{align*}
Then,  the achievable rate can be numerically evaluated for the specified path.
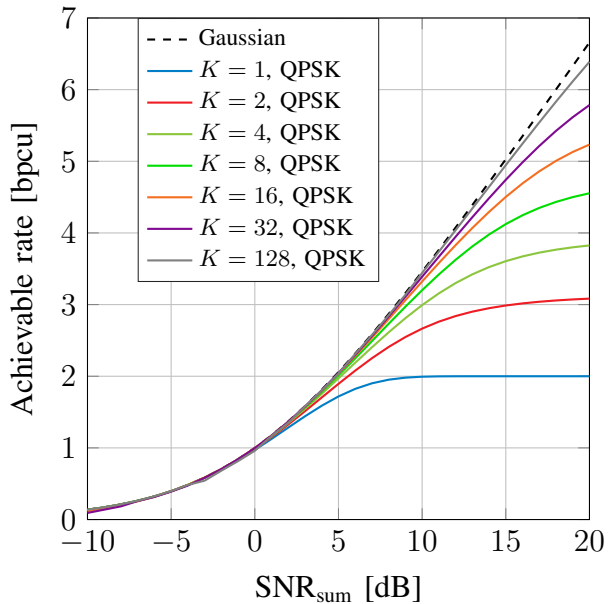
\begin{figure}[tbp]
\centering
\begin{tikzpicture}

\pgfplotsset{compat=1.12}
\begin{axis}[
        width=.5\linewidth,
	    height=.5\linewidth,
        xmajorgrids,
        yminorticks=true,
        ymajorgrids,
        yminorgrids,
        legend pos=south west,        
        legend style={at={(0.1,1)},
      anchor=north west, legend columns=1, legend cell align=left,align=left,draw=white!15!black, font=\footnotesize},
        xlabel={$\textrm{SNR}_{\textrm{sum}}$ [dB]},
        ylabel={Achievable rate [bpcu]},
        mark size=1.5pt,
        xmin=-10,
        xmax=20,
        ymin=0,
        ymax=7
    ]	
	\addplot[color=black, thick,dashed] table {tikz/data/AchievableRate_7.dat};
	\addlegendentry{Gaussian}
      \addplot[color=mittelblau, thick] table {tikz/data/AchievableRate_8.dat};
	\addlegendentry{$K=1$, QPSK}
	\addplot[color= rot, thick] table {tikz/data/AchievableRate_6.dat};
	\addlegendentry{$K=2$, QPSK}
      \addplot[color= apfelgruen, thick] table {tikz/data/AchievableRate_5.dat};
	\addlegendentry{$K=4$, QPSK}
	\addplot[color=gelb, thick] table {tikz/data/AchievableRate_4.dat};
	\addlegendentry{$K=8$, QPSK}	
	\addplot[color=orange, thick] table {tikz/data/AchievableRate_3.dat};
	\addlegendentry{$K=16$, QPSK}	
      \addplot[color=lila, thick] table {tikz/data/AchievableRate_2.dat};
	\addlegendentry{$K=32$, QPSK}	
       \addplot[color=mittelgrau, thick] table {tikz/data/AchievableRate_1.dat};
	\addlegendentry{$K=128$, QPSK}		
 
\end{axis}

\end{tikzpicture}
\vspace*{-0.5cm}
  \caption{Achievable rates of multiuser IDMA with matching codes and QPSK modulation; all users are assumed to have the same power, modulation and coding scheme; For fair comparison, the multi-user SNR $\mathrm{SNR}_{\mathrm{sum}}=KP/\sigma^{2}$  is used as abscissa.}
  \label{Fig:QPSK_Rate}
  \vspace*{-0.5cm}
\end{figure}
We compare the achievable rates using QPSK with different number of users in Fig.~\ref{Fig:QPSK_Rate} by numerically solving the integral \eqref{eq:UserRateIntQPSK}. Clearly, the loss to Gaussian capacity, due to finite modulation, can be made arbitrarily small by imposing a larger number of users or date layers (which may belong to a same user) into the system. Although we assumed equal-power and equal-rate for simplicity, the achievable rates analysis can be extended to other general cases straightforwardly.   

We will provide code matching examples for a three user case based
on  QPSK signaling in Sec. \ref{sec:Results}. The achievable rates can also be found by the
density evolution (DE) method, which are very close to the Gaussian
capacity.

\begin{figure*}[htbp]
\begin{centering}
\includegraphics[width=1.0\columnwidth]{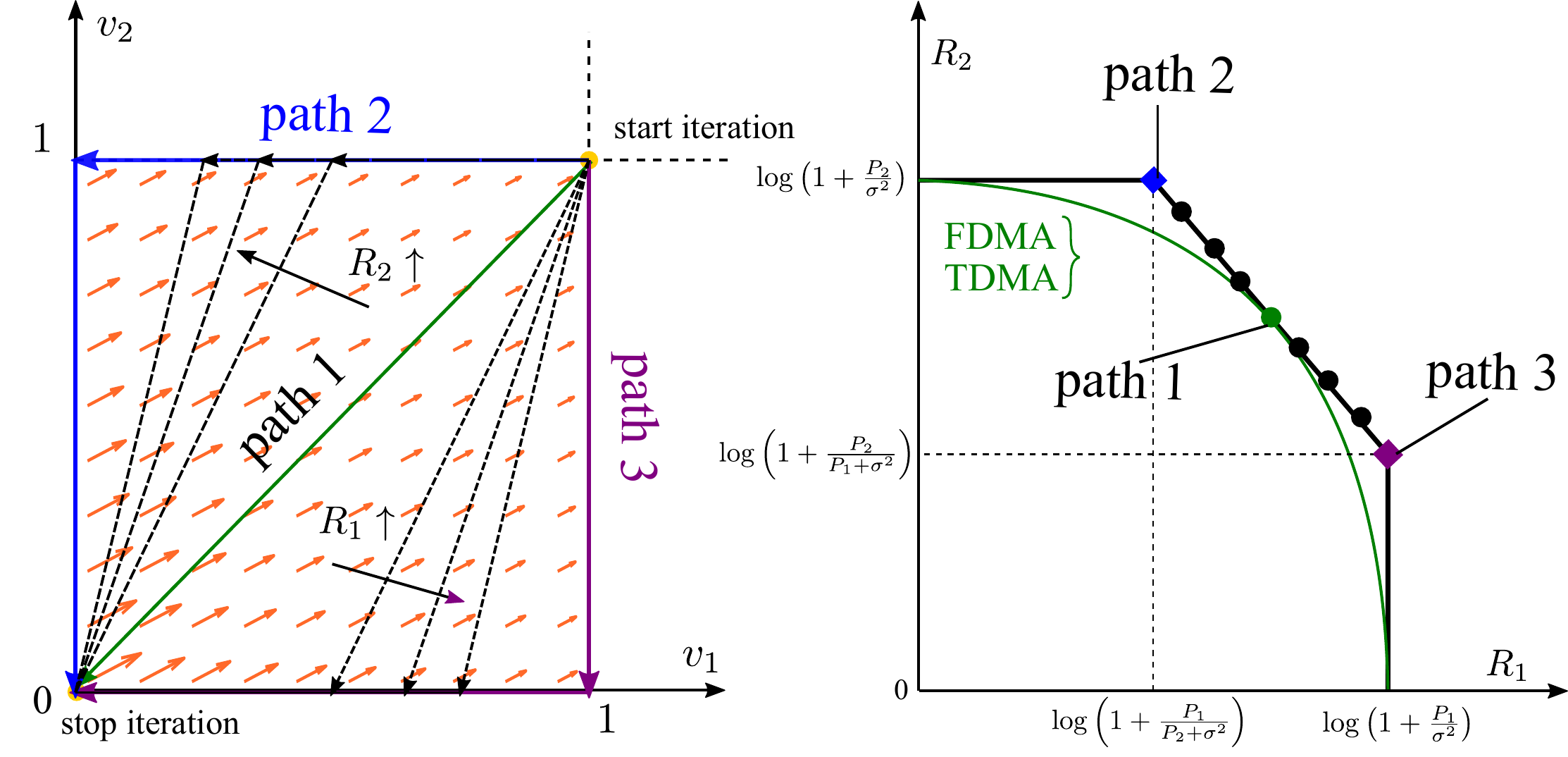}
\par\end{centering}
\vspace*{-0.5cm}
\caption{Illustration (exemplary for two users) of different integration paths
achieving different rate pairs $\left(R_{1},R_{2}\right)$; the arrows
in the left figure illustrate the two-dimensional MSE vector field;
the achieved rate pairs are marked in the right figure for the corresponding
paths; the dashed lines denote paths achieving rate pairs moving (from the green dot) toward the corresponding SIC corner points.\label{fig:Illustration-of-different}}
\vspace*{-0.5cm}
\end{figure*}

\subsection{Path vs rate tuples}\label{subsec:PathExam}

Consider a simple two-user case, i.e., $K=2$. Fig.~\ref{fig:Illustration-of-different}
illustrates some special paths and their corresponding achievable
rate tuple (or rate pair here).
The simplest path is a straight line between the starting point $\boldsymbol{v}\left(t_{0}\right)=\mathbf{1}$
and the stop point $\boldsymbol{v}\left(t_{\infty}\right)=\mathbf{0}$,
denoted by \textit{path 1}. It is straightforward to obtain
\[
R_{k}=\frac{g_{k}}{\mathbf{g}^{T}\mathbf{1}}\mathrm{log}\left(1+\frac{\mathbf{g}^{T}\mathbf{1}}{\sigma^{2}}\right),\,\forall k=1,2,\ldots,K.
\]
In this case, the achievable rate of each user is proportional to
the received signal power strength $g_{k}$. For the two-user case, this
rate tuple coincides with the point where TDMA/FDMA achieves the sum-rate
capacity (see green dot in Fig.~\ref{fig:Illustration-of-different}).
In \textit{path 1, }it satisfies
\[
v_{1}\left(t\right)=v_{2}\left(t\right)=\cdots=v_{K}\left(t\right)=v\left(t\right),\,\forall\,t.
\]
The matching code for $k$th user shall have the following MSE characteristic
function
\[
v_{k}=\begin{cases}
1 & \rho_{k}\leq\rho_{k,\mathrm{min}}\\
\frac{1}{\mathbf{g}^{T}\mathbf{1}-g_{k}}\cdot\left(\frac{1}{\rho_{k}}-\sigma^{2}\right) & \rho_{k,\mathrm{min}}\leq\rho_{k}\leq\rho_{k,\mathrm{max}}\\
0 & \rho_{k}\geq\rho_{k,\mathrm{max}}
\end{cases}.
\]

For \textit{path 2 }and\textit{ path 3 }which are comprised of $K$
segments and each segment has merely value change (from $v_{l}=1$
to $v_{l}=0$) in one particular direction $v_{l}$, i.e., within
the segment $\frac{dv_{l}}{dt}\ne0$ and $\frac{dv_{k}}{dt}=0,\forall k\ne l$.
Depending on the order of the segments, there exist $K!$ such paths,
which constitute the $K!$ SIC corner points of MAC capacity region.
The user rate can be written as
\begin{align*}
R_{k}=\mathrm{log}
\Bigg(
1+\dfrac{ g_{k}}{{\sum\limits_{l=\pi\left(k\right)+1}^{K}g_{l}} +\sigma^{2} }
\Bigg),\,\forall k=1,2,\ldots,K,
\end{align*}
where $\pi\left(k\right)=k'$ denotes the permutation of user order
with $1\leq\pi\left(k\right)\leq K$ and $\pi\left(k\right)\ne\pi\left(k'\right),\forall k\ne k'$.
The corresponding decoding functions are given by
\[
v_{k}=\psi_{k}\left(\rho_{k}\right)=\begin{cases}
1 & \rho_{k}<\rho_{k,\mathrm{SIC}}\\
0 & \rho_{k}\geq\rho_{k,\mathrm{SIC}}
\end{cases}
\]
where
\[
\rho_{k,\mathrm{SIC}}=\frac{g_{k}}{{\displaystyle \sum_{l=\pi\left(k\right)+1}^{K}}g_{\pi\left(l\right)}+\sigma^{2}}
\]
are the decoding thresholds. The decoding functions are step functions
with sharp transitions at corresponding threshold SNRs $\rho_{k,\mathrm{SIC}}$.
This type of decoding functions may pose difficulties for practical code designs, compared to that with smooth transitions.

\subsection{Achievable rate region}\label{subsec:RateRegion}

To achieve an arbitrary point of the MAC capacity region, other paths shall be found. In the following theorem, we show that the entire MAC capacity region can be achieved by proving the existence of a dedicated path achieving an arbitrary point within the capacity region. Examples for constructing a dedicated path achieving a feasible rate tuple are provided in \textit{case 2} of Sec.~\ref{sec:ESE-function}.
\begin{thm}
\label{thm:IDMA-achieves-everywhere}Under the assumptions in Theorem 1, IDMA with GA-based MUD achieves every rate tuple
in the $K$-user MAC capacity region $\mathcal{C\left(\mathit{K}\right)}$.
Given a feasible target rate tuple $\mathbf{R}=\left[R_{1},R_{2},\cdots,R_{K}\right]\in\mathcal{C\left(\mathit{K}\right)}$,
there exists at least one path defined by $\boldsymbol{v}_{R}\left(t\right):\boldsymbol{v}_{s}=\mathbf{1}\rightarrow\boldsymbol{v}_{e}=\mathbf{0}$
which achieves $\mathbf{R}$.
\end{thm}
\begin{IEEEproof}
See Appendix B.
\end{IEEEproof}
\textit{Remark: }It is easy to prove that there exists a unique path
for each of the $K!$ SIC corner points and the decoding functions
shall be step functions. For other rate tuples, it can be verified
that there exist many different paths achieving that rate tuple. The
choice of the integration path poses varying degrees of difficulty for the
design of matching codes. Thus, the design of an appropriate integration
path could be an extra degree of freedom for code design.


\subsection{Gaussian Approximation}\label{subsec:GA}
We provide in this section numerical evidence showing that the GA used in our achievable rate analysis is accurate enough for addressing the behavior of
a practical iterative IDMA multiuser demodulator and decoder. The technique we used to track the probability density function (PDF) of the exchanged messages during the iterative processing is discretized density evolution (DDE) proposed in \cite{SaeLDPCDDE}.

The GA is arguably true for large number of users (central limit theorem) and/or noise-limited scenarios (the noise density rather the multiple access interference governs the iterative process). We verified the GA through DDE for these cases (results omitted). Instead, we show results for the following example with a few number of users operating at relatively high SNR, since the GA becomes skeptical in these cases.    

\begin{figure}[htbp]
\begin{center}
\begin{tikzpicture}

\pgfplotsset{compat=1.12}
\begin{axis}[
        width=.5\linewidth,
	    height=.5\linewidth,
        xmajorgrids,
        yminorticks=true,
        ymajorgrids,
        yminorgrids,
        legend pos=south west,        
        legend style={at={(0.6,1)},
      anchor=north west, legend columns=1, legend cell align=left,align=left,draw=white!15!black, font=\footnotesize},
        xlabel={LLR at VND},
        ylabel={prob. density},
        mark size=1.5pt,
        xmin=-10,
        xmax=25,
        ymin=0,
        ymax=0.015
    ]	

      \addplot[color=mittelblau, thick] table {tikz/data/LLR_dist8.dat};
	\addlegendentry{iteration 1}
	\addplot[color= rot, thick] table {tikz/data/LLR_dist7.dat};
	\addlegendentry{iteration 2}
      \addplot[color= apfelgruen, thick] table {tikz/data/LLR_dist6.dat};
	\addlegendentry{iteration 3}
	\addplot[color=gelb, thick] table {tikz/data/LLR_dist5.dat};
	\addlegendentry{iteration 4}	
	\addplot[color=orange, thick] table {tikz/data/LLR_dist4.dat};
	\addlegendentry{iteration 5}	
      \addplot[color=lila, thick] table {tikz/data/LLR_dist3.dat};
	\addlegendentry{iteration 6}	
       \addplot[color=mittelgrau, thick] table {tikz/data/LLR_dist2.dat};
	\addlegendentry{iteration 7}	
 
\end{axis}

\end{tikzpicture}
\end{center}
\vspace*{-1cm}
\caption{LLR distribution at the VND during the iterative multiuser detection and decoding model with $K=4$ users at the SNR of 20~dB; the symbol $x=+1$ is assumed to be transmitted for the user under test.}
\label{fig:LLR_GA}
\vspace*{-0.3cm}
\end{figure}
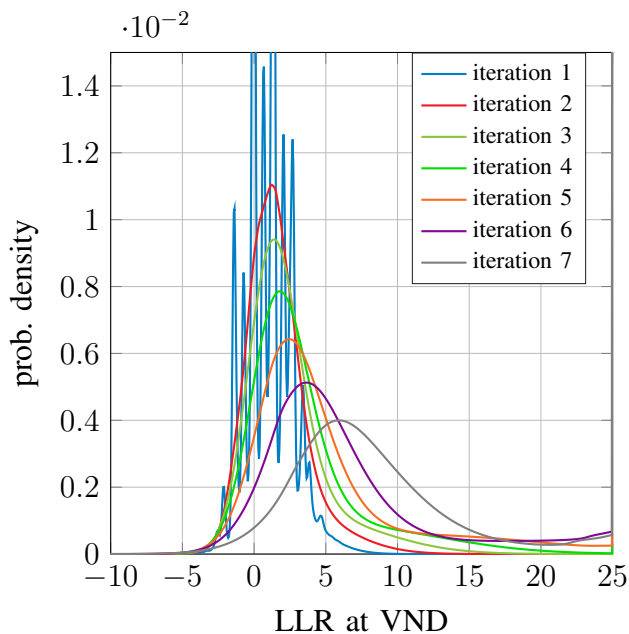

\textit{Example: }The number of users is set to $K=4$, each with the same power $P=\frac{1}{4}$ and BPSK modulation. The multiuser SNR is $-10\log\sigma^{2}=$20~dB.
An LDPC code with the variable node degree profile (from ``edge'' perspective) $0.5231\lambda^{1} + 0.3187\lambda^{2}+ 0.1582\lambda^{11}$ and the check node degree $\eta^{2}$ is used for each user.
The PDF of Log-likelihood ratio (LLR) at the output of the LDPC variable node decoder (VND) is tracked using DDE with 10 bits and shown in Fig.~\ref{fig:LLR_GA} for the first 8 iterations. 
Clearly, the interference plus noise does not resemble a Gaussian density at the first iteration. As the consequence of the soft interference cancellation, the density at VND becomes more Gaussian-like as the iteration proceeds. Similar trend can be observed also at the output of ESE and CND (results not shown). Surprisingly, the GA is quite accurate even for a few number of users operating at high SNR regime.

\section{MU-MIMO Channel}\label{sec:MIMO}

Assume that the $k$th transmitter has $N_{t,k}$ antennas and the receiver has
$N_{R}$ antennas respectively; then, the received signal can be written as
\begin{equation}
\mathbf{y}={\displaystyle \sum_{k=1}^{K}}\sqrt{P_{k}}\mathbf{H}_{k}\mathbf{x}_{k}+\mathbf{n}\tag{11}\label{eq:MIMOsysM}
\end{equation}
where $\mathbf{H}_{k}$ is the channel of the $k$th user, $\mathbf{n}$
denotes the uncorrelated noise $\mathrm{E}\left[\mathbf{n}\mathbf{n}^{H}\right]=\sigma^{2}\mathbf{I}$.
In this case, the ESE module is replaced by an iterative linear MMSE (LMMSE)
receiver \cite[eqn. (4a)]{XYuanITI14}. Under the LMMSE-based ESE, the SNR of
user $k$ can be written as \cite{YuanMIMOLMMSESINR}
\begin{equation}
\rho_{k}=\frac{{\displaystyle \sum_{i=1}^{N_{t,k}}}\mathbf{h}_{k,i}^{H}\mathbf{R}^{-1}\mathbf{h}_{k,i}}{1-{\displaystyle v_{k}\sum_{i=1}^{N_{t,k}}}\mathbf{h}_{k,i}^{H}\mathbf{R}^{-1}\mathbf{h}_{k,i}}\tag{12}
\end{equation}
where $\mathbf{h}_{k,i}$ denotes the $i$th column of the $k$th
user's channel matrix $\mathbf{H}_{k}$ and
\[
\mathbf{R}=\sigma_{n}^{2}\mathbf{I}+\mathbf{HVH}^{H}
\]
with $\mathbf{V}=\mathrm{diag}\left(P_{1}v_{1},P_{2}v_{2},\cdots,P_{K}v_{K}\right)$
and $\mathbf{H}$ being the concatenated channels of all users. Following
a similar approach in Appendix A, we obtain with the matching condition
in \eqref{eq:MatchCondLine} the user rate $R_k$ as
\[
R_{k}=\left[-\int{\displaystyle \sum_{i=1}^{N_{t,k}}}\mathbf{h}_{k,i}^{H}\mathbf{R}^{-1}\mathbf{h}_{k,i}dv_{k}\right]_{v_{k}=1}^{v_{k}=0}
\]
Therefore, the sum-rate can be obtained as
\begin{align}
R_{\mathrm{sum}} & ={\displaystyle \sum_{i=1}^{K}R_{i}}=-\int_{\mathbf{v}=\mathbf{1}}^{\mathbf{v}=\mathbf{0}}\nabla\mathrm{log\,det}\left[\mathbf{R}\right]d\mathbf{v}\nonumber \\
 & =\mathrm{log\,det}\left[\mathbf{I}+\frac{1}{\sigma_{n}^{2}}\mathbf{H}^{H}\mathbf{P}\mathbf{H}\right]\tag{13}
\end{align}
where $\mathbf{P}=\mathrm{diag}\left(P_{1},P_{2},\cdots,P_{K}\right)$.
Path independence follows from the condition
\[
\frac{\partial}{\partial v_{k}}\mathrm{log}\mathrm{\,det}\left[\mathbf{R}\right]=\mathrm{trace}\left[\mathbf{R}^{-1}\mathbf{H}_{k}\mathbf{H}_{k}^{H}\right]={\displaystyle \sum_{i=1}^{N_{t,k}}}\mathbf{h}_{k,i}^{H}\mathbf{R}^{-1}\mathbf{h}_{k,i}
\]
with Jacobi's formula.

\section{Results }\label{sec:Results}
The code design for multi-user can be complicated \cite{SharifiTWC1662userLDPC,ABSTWCL182userLDPC, MIMONOMASonLiu,GSongIT16}.
For simplicity, we consider a single-input single-output (SISO) setup. We assume that the power levels $g_i = P_{i}|h_i|^{2} (i=1,\cdots,K)$ are constant in our code design. Thus, the multi-user SNR is defined as
\begin{equation}\label{eq:sum_snr}
\mathrm{SNR}_{\mathrm{sum}}= \frac{\sum_{i=1}^{K}g_i}{\sigma^2}. \tag{14}
\end{equation}

We consider $K=3$ users with the power distribution $\bm{g} = [g_1, g_2, g_3]^T = [\frac{1}{7}, \frac{2}{7}, \frac{4}{7}]^T$ and we target the sum-rate $R_{\mathrm{sum}} = R_1 + R_2 + R_3 = 1$~bpcu as an example.
Theoretically, this sum-rate is attainable at the noise variance $\sigma^2 = 1$.
Furthermore, the capacity region (more precisely, the \textit{dominant face} which maximizes the sum-rate) with Gaussian alphabets is given by
\begin{align}\label{eq:rateConstraint}
  0.1069 \leq R_1 \leq & \log_2\left( 1 + \frac{g_1}{\sigma^2} \right) = 0.1926,  \nonumber \\
  0.2224 \leq R_2 \leq & \log_2\left( 1 + \frac{g_2}{\sigma^2} \right) = 0.3626,  \nonumber \\
  0.4854 \leq R_3 \leq & \log_2\left( 1 + \frac{g_3}{\sigma^2} \right) = 0.6521,  \nonumber \\
  R_1 + R_2 \leq & \log_2\left( 1 + \frac{g_1 + g_2}{\sigma^2} \right) = 0.5145, \nonumber \\
  R_1 + R_3 \leq & \log_2\left( 1 + \frac{g_1 + g_3}{\sigma^2} \right) = 0.7776, \nonumber \\
  R_2 + R_3 \leq & \log_2\left( 1 + \frac{g_2 + g_3}{\sigma^2} \right) = 0.8931. \nonumber
\end{align}

\subsection{ESE Functions}\label{sec:ESE-function}
According to the matching condition in $(6)$, for the design of capacity-achieving codes, the ESE transfer functions $\bm{\rho}(t) = \phi(\bm{v}(t))$ shall be determined. For this, we specify the $K$-dimensional decoding path $\bm{v}(t)$.


As the path independence property of Theorem \ref{thm:IDMA-is-capacity-achieving}, we can constraint $\bm{v}(t)$ to be a piece-wise linear path with $n$ segments starting from the point $\bm{v}(t=0)=\bm{x}_0=\bm{1}$, crossing the intermediate points $\bm{v}(t=i)=\bm{x}_i=[x_{i,1},\cdots,x_{i,K}]^T, i=1,2,\cdots,n-1$, and terminating at the point $\bm{v}(t=n)=\bm{x}_n=\bm{0}$, where $\bm{x}_i \neq \bm{x}_j, \forall i \neq j$ and for practical decoding
\begin{equation}\label{eq:pathConst} 
1 \geq x_{1,k} \geq x_{2,k} \geq \cdots \geq x_{n-1,k} \geq 0 \, \forall k \tag{15}
\end{equation}
shall apply.
Therefore, the path can be expressed in  a vector form as
\begin{equation}\label{eq:path}
 \bm{v}(t) = \bm{x}_i - (\bm{x}_i - \bm{x}_{i+1}) \cdot (t-i), t \in [i, i+1] \tag{16}
\end{equation}
for $i = 0,1,2,\cdots,n-1$. With the specified path, the ESE transfer function for user $k$ can be computed as
\begin{align}\label{eq:path_transfer}
\rho_k &=\phi_{k}\left(v_k \right)= \frac{g_k}{\bm{g}^T\bm{v}(t) - g_k v_k(t) + \sigma^2} \nonumber \\
       &= \frac{g_k}{\bm{g}^T\left[\bm{x}_i - (\bm{x}_i - \bm{x}_{i+1}) \cdot \frac{x_{i,k}-v_k}{x_{i,k}-x_{i+1,k}}\right] - g_k v_k + \sigma^2}, \nonumber \\
       \!=&\! \frac{\!g_k}{\!\sum\limits_{\!k' \neq k}^{\!K} \! {g}_{k'} \! \left( \!\frac{\!{x}_{\!i,k'} - {x}_{i+1,k'}}{x_{i,k}-x_{i+1,k}}v_k + \frac{x_{i,k}{x}_{i+1,k'}-x_{i+1,k}{x}_{i,k'}}{x_{i,k}-x_{i+1,k}} \! \right) + \sigma^2}, \nonumber \\
       & v_k \in [x_{i+1,k}, x_{i,k}] \mbox{~for~} i = 0,1,2,\cdots,n-1. \tag{17}
\end{align}
Note that when $x_{i,k} =  x_{i+1,k}$, the above function is not valid. Actually, $\rho_k$ is a vertical line from $\frac{g_k}{\bm{g}^T\bm{x}_i - g_k x_{i,k} + \sigma^2}$ to $\frac{g_k}{\bm{g}^T\bm{x}_{i+1} - g_k x_{i+1,k} + \sigma^2}$ with $v_k = x_{i,k}$. Substituting \eqref{eq:path} into \eqref{eq:UserRateGaussian}, we obtain the user rate $R_k$
\begin{align}\label{eq:rate}
R_k &= -\int_{v_k(t)=1}^{v_k(t)=0} \frac{g_k}{\bm{g}^T\bm{v}(t)+\sigma^2} dv_k(t) \nonumber \\
   =& \sum\limits_{i=0}^{n-1} \frac{g_k(x_{i,k}-x_{i+1,k})}{\bm{g}^T(\bm{x}_i - \bm{x}_{i+1})} \log \frac{\bm{g}^T\bm{x}_{i}+\sigma^2}{\bm{g}^T\bm{x}_{i+1}+\sigma^2}. \tag{18}
\end{align}

To verify the path independence property of Theorems \ref{thm:IDMA-is-capacity-achieving} and \ref{thm:IDMA-achieves-everywhere}, we consider three different paths for \eqref{eq:path_transfer} and evaluate the system performance and achievable rates via density evolution and bit error rate (BER) simulations.
\begin{figure*}[htbp]
  \includegraphics[width=\textwidth]{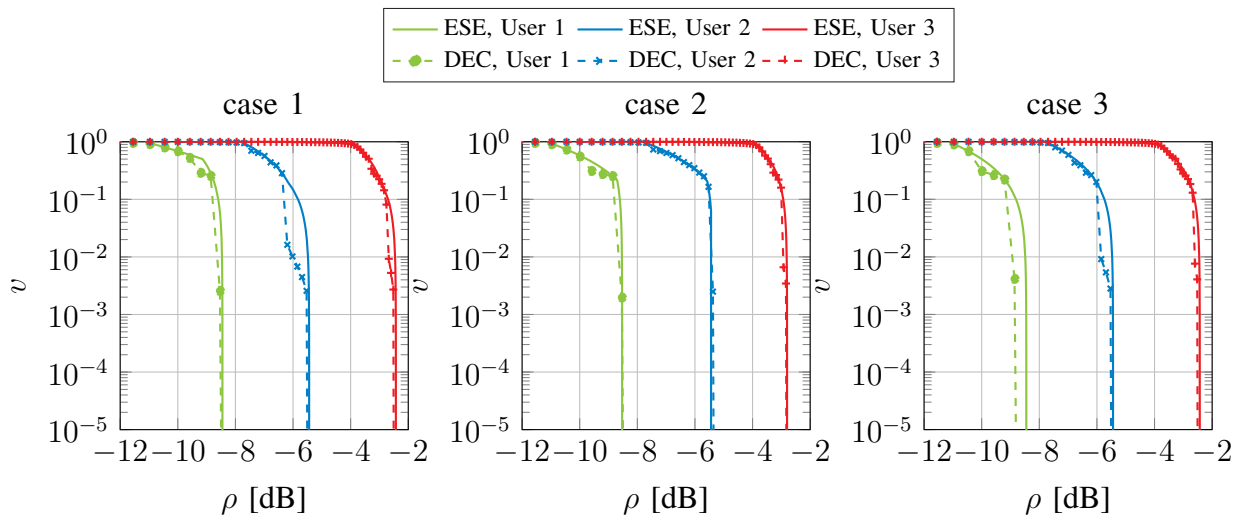}
  \vspace*{-0.3cm}
  \caption{ESE transfer function and the matching LDPC code transfer function for three different paths; the x-axis denotes the SNR of ESE and the y-axis denotes the MSE of the feedback from channel decoder; three users with QPSK and the power distribution  $\bm{g} = [g_1, g_2, g_3]^T = [\frac{1}{7}, \frac{2}{7}, \frac{4}{7}]^T$ are considered.}
  \label{Fig:ESE_DEC_transfer}
\end{figure*}

\subsubsection{Case 1}
We do not specify any intermediate point $\{\bm{x}_i\}$, i.e., $n = 1$. The path is a straight line between the starting point $\bm{v}(t=0)=\bm{1}$ and the stop point $\bm{v}(t=\infty)=\bm{0}$, as discussed in Sec. \ref{subsec:PathExam}. The ESE function for user $k$ is given by
      \begin{equation}\label{eq:Case1_ESE_transfer}
        \rho_{k} = \frac{g_k}{(\bm{g}^T\bm{1}-g_k) v_k + \sigma^2}, v_k \in [0, 1]. \tag{19}
      \end{equation}
      The rate for user $k$ is proportional to its power $g_k$, i.e., $R_k = \frac{g_k}{\bm{g}^T\bm{1}}R_{\mathrm{sum}}$. Thus, the corresponding rate tuple is $(R_1, R_2, R_3) = (\frac{1}{7}, \frac{2}{7}, \frac{4}{7})$. The transfer functions in (\ref{eq:Case1_ESE_transfer}) are depicted in the left most sub-figure in Fig.~\ref{Fig:ESE_DEC_transfer} for the three users, respectively.

\subsubsection{Case 2}
We construct a dedicated path to achieve an arbitrarily chosen rate tuple in the MAC region, e.g., $(R_1, R_2, R_3) = (0.15, 0.3, 0.55)$. To find a dedicated path, we search for $\{\bm{x}_i\}$ by solving $K$ non-linear equations given by \eqref{eq:rate}. Then, the ESE transfer functions can be obtained by substituting $\{\bm{x}_i\}$ into (\ref{eq:path_transfer}). If $n>2$, there are $K(n-1)$ unknown variables $\{x_{i,k}\}$, which is larger than $K$. This potentially result in multiple solutions. It is noteworthy to mention that the variables $\{x_{i,k}\}$ are bounded in $[0,1]$ and shall satisfy \eqref{eq:pathConst}. We may fix some unknown variables $\{x_{i,k}\}$ and solve the $K$ non-linear equations given by \eqref{eq:rate} to obtain remaining unknown variables. Usually, we can fix $K(n-2)$ unknown variables and have feasible solution for the remaining $K$ unknown variables. Here, we consider a $3$-segment path having intermediate points
        \begin{equation}\label{eq:intermediate_points}
          \bm{x}_{1} = [x_{1,1}, x_{1,2}, 0]^T \mbox{~and~} \bm{x}_{2} = [0, x_{2,2}, 0]^T. \tag{20}
        \end{equation}
        Substituting (\ref{eq:intermediate_points}) into (\ref{eq:rate}), we obtain
        \begin{small}
        \begin{align}
          0.15 &=  \frac{g_1(1-x_{1,1})}{1-{g}_1{x}_{1,1}-{g}_2{x}_{1,2}} \log_2 \frac{2}{{g}_1{x}_{1,1}+{g}_2{x}_{1,2}+1} \nonumber \\
               & + \frac{g_1{x}_{1,1}}{g_1{x}_{1,1} + g_2(x_{1,2}-x_{2,2})} \log_2 \frac{g_1 x_{1,1} + g_2 x_{1,2}+1}{g_2 x_{2,2}+1}, \nonumber \\
          0.3  &=  \frac{g_2(1-x_{1,2})}{1-{g}_1{x}_{1,1}-{g}_2{x}_{1,2}} \log_2 \frac{2}{{g}_1{x}_{1,1}+{g}_2{x}_{1,2}+1} \nonumber \\
               & + \frac{g_2(x_{1,2}-x_{2,2})}{g_1{x}_{1,1} + g_2(x_{1,2}-x_{2,2})} \log_2 \frac{g_1 x_{1,1} + g_2 x_{1,2}+1}{g_2 x_{2,2}+1} \nonumber \\
               & + \log_2 \left( g_2 x_{2,2}+1\right), \nonumber \\
          0.55 &= \frac{g_{3}}{1-{g}_1{x}_{1,1}-{g}_2{x}_{1,2}} \log_2 \frac{2}{{g}_1{x}_{1,1}+{g}_2{x}_{1,2}+1}. \nonumber \\
          \tag{21} \label{eq:Case2_Equations}
        \end{align}
        \end{small}
        Solving (\ref{eq:Case2_Equations}), we obtain one  feasible solution given by $x_{1,1} = 0.2145, x_{1,2} = 0.2056, x_{2,2} = 0.0618$. Substituting the solution into (\ref{eq:path_transfer}), we can obtain the ESE functions.
        These transfer functions are depicted in the middle sub-figure of Fig.~\ref{Fig:ESE_DEC_transfer} for the three users, respectively.

\subsubsection{Case 3}
We randomly choose the intermediate points $\{\bm{x}_i\}$. Then, substituting the points into (\ref{eq:path_transfer}), we obtain the ESE transfer functions and subsequently compute the rate for each user using (\ref{eq:rate}).

Here, we consider a piece-wise linear path with $2$ segments by specifying an intermediate point arbitrarily, e.g., $\bm{x}_1 = [0.5, 0.2, 0.2]$. Substituting $\bm{x}_1$ into (\ref{eq:path_transfer}), we have the ESE functions as
\begin{eqnarray}\label{eq:Case3_transfer}
  \rho_1 \!&=\! \left\{\! \begin{array}{ll}
                           \frac{g_1}{(g_2+g_3)0.4v_1+\sigma^2}, & 0 \leq v_1 \leq 0.5, \\
                           \frac{g_1}{g_2(1.6v_1-0.6)+g_3(1.6v_1-0.6)+\sigma^2}, & 0.5 \leq v_1 \leq 1, \\
                         \end{array}
                \right. \nonumber \\
  \rho_2 \!&=\! \left\{\! \begin{array}{ll}
                           \frac{g_2}{g_1 2.5v_2+g_3 v_2+\sigma^2}, & 0 \leq v_2 \leq 0.2, \\
                           \frac{g_2}{g_1(0.625v_2+0.375)+g_3v_2+\sigma^2}, & 0.2 \leq v_2 \leq 1, \\
                         \end{array}
               \right. \nonumber \\
  \rho_3 \!&=\! \left\{\! \begin{array}{ll}
                           \frac{g_3}{g_1 2.5v_3+g_2v_3+\sigma^2}, & 0 \leq v_3 \leq 0.2, \\
                           \frac{g_3}{g_1(0.625v_3+0.375)+g_2v_3+\sigma^2}, & 0.2 \leq v_3 \leq 1. \\
                         \end{array}
               \right. \nonumber
\end{eqnarray}
These transfer functions are depicted in the right most sub-figure of Fig.~\ref{Fig:ESE_DEC_transfer} for the three users, respectively.
The achievable rates for each user,  given in (\ref{eq:Case3_rate}),  can be obtained by  substituting $\bm{x}_1$ into (\ref{eq:rate}), see Tab.~\ref{tab:OptResult}.
\begin{figure*}[h]
\begin{align}
  R_1 &=&  \frac{g_1(1 - 0.5)}{g_1(1-0.5)+g_2(1-0.2)+g_3(1-0.2)} \log_2\left( \frac{1 + 1}{g_1 \cdot 0.5 + g_2 \cdot 0.2 + g_3 \cdot 0.2 + 1} \right) \nonumber \\
      && + \frac{g_1(0.5 - 0)}{g_1(0.5-0)+g_2(0.2-0)+g_3(0.2-0)} \log_2\left( {g_1 \cdot 0.5 + g_2 \cdot 0.2 + g_3 \cdot 0.2 + 1} \right) = 0.157, \nonumber \\
  R_2 &=& \frac{g_2(1 - 0.2)}{g_1(1-0.5)+g_2(1-0.2)+g_3(1-0.2)} \log_2\left( \frac{1 + 1}{g_1 \cdot 0.5 + g_2 \cdot 0.2 + g_3 \cdot 0.2 + 1} \right) \nonumber \\
      && + \frac{g_2(0.2 - 0)}{g_1(0.5-0)+g_2(0.2-0)+g_3(0.2-0)} \log_2\left( \frac{g_1 \cdot 0.5 + g_2 \cdot 0.2 + g_3 \cdot 0.2 + 1}{0 + 1} \right) = 0.281, \nonumber \\
  R_3 &=&  \frac{g_3(1 - 0.2)}{g_1(1-0.5)+g_2(1-0.2)+g_3(1-0.2)} \log_2\left( \frac{1 + 1}{g_1 \cdot 0.5 + g_2 \cdot 0.2 + g_3 \cdot 0.2 + 1} \right) \nonumber \\
      && + \frac{g_3(0.2 - 0)}{g_1(0.5-0)+g_2(0.2-0)+g_3(0.2-0)} \log_2\left( \frac{g_1 \cdot 0.5 + g_2 \cdot 0.2 + g_3 \cdot 0.2 + 1}{0 + 1} \right) = 0.562.
\tag{22}\label{eq:Case3_rate}
\end{align}
\hrulefill
\end{figure*}
\begin{figure*}[t]
\begin{equation}\label{eq:evo}
I_{E,V} = \sum\limits_{i=1}^{d_{v,\max}}\!\lambda_i \cdot J \! \left( \sqrt{(i-1)\left[ J^{-1}\left( 1 -\! \sum\limits_{j=1}^{d_{c,\max}}\eta_j \cdot J\left( \sqrt{j-1} \cdot J^{-1}\left(1-I_{E,V} \right) \right) \right) \right]^2 + 4\rho} \right), \tag{27}
\end{equation}
\hrulefill
\end{figure*}

\subsection{LDPC code optimization}\label{sec:LDPCopt}
\newcommand{\tabincell}[2]{\begin{tabular}{@{}#1@{}}#2\end{tabular}}
\begin{table*}[htbp]
\caption{Code Optimization Results for Three Cases}
\centering\label{tab:OptResult}
\begin{tabular}{|c||c|c|c||c|c|c|}
\hline
Case                        & \multicolumn{3}{c||}{Case 1, $R_{\mathrm{sum}}=1$}                        & \multicolumn{3}{c|}{Case 2, $R_{\mathrm{sum}}=1$}  \\
\hline
User                        & {User 1}        & {User 2}        & {User 3}        & {User 1}        & {User 2}        & {User 3}    \\
\hline
Power                       & {$\frac{1}{7}$} & {$\frac{2}{7}$} & {$\frac{4}{7}$} & {$\frac{1}{7}$} & {$\frac{2}{7}$} & {$\frac{4}{7}$} \\
\hline
Path                        & \multicolumn{3}{c||}{$[1,1,1]\rightarrow[0,0,0]$}   & \multicolumn{3}{c|}{\tabincell{c}{$[1,1,1]\rightarrow[0.2145,0.2056,0]$ \\ $\rightarrow[0,0.0618,0]\rightarrow[0,0,0]$}}  \\
\hline
\tabincell{c}{Target rate} & {$0.1429$}      & {$0.2857$}      & {$0.5714$}      & {$0.15$}        & {$0.30$}        & {$0.55$} \\
\hline
{\tabincell{c}{Check edge\\distribution}} & $\eta_3=1$ & $\eta_4=1$ & $\eta_5=1$ & $\eta_3=1$      & $\eta_4=1$      & $\eta_5=1$ \\
\hline
{\tabincell{c}{variable degree\\set $\{d_{v}\}$}} & \multicolumn{3}{c||}{\{2:1:30, 35:5:50, 60:10:100\}} & \multicolumn{3}{c|}{\{2:1:30, 35:5:50\}} \\
\hline
                            &$\lambda_2,0.5239$    &$\lambda_2,0.3770$    &$\lambda_2,0.3293$    &$\lambda_2,0.5234$    &$\lambda_2,0.3779$    &$\lambda_2,0.3218$     \\
 Optimized                  &$\lambda_3,0.2140$    &$\lambda_3,0.2168$    &$\lambda_3,0.2351$    &$\lambda_3,0.2292$    &$\lambda_3,0.2290$    &$\lambda_3,0.2273$      \\
 variable                   &$\lambda_7,0.1627$    &$\lambda_7,0.0719$    &$\lambda_8,0.2500$    &$\lambda_7,0.0896$    &$\lambda_7,0.1431$    &$\lambda_7,0.0879$       \\
 edge                       &$\lambda_{30},0.0685$ &$\lambda_8,0.1577$    &$\lambda_{21},0.0654$ &$\lambda_8,0.0590$    &$\lambda_8,0.0580$    &$\lambda_8,0.1654$        \\
 distribution               &$\lambda_{35},0.0309$ &$\lambda_{40},0.1237$ &$\lambda_{22},0.0014$ &$\lambda_{30},0.0708$ &$\lambda_{50},0.1920$ &$\lambda_{20},0.0501$  \\
                            &                      &$\lambda_{100},0.0529$&$\lambda_{45},0.0258$ &$\lambda_{35},0.0280$ &                      &$\lambda_{21},0.0335$  \\
                            &                      &                      &$\lambda_{50},0.0930$ &                      &                      &$\lambda_{50},0.1140$                       \\
\hline
\tabincell{c}{Optimized\\rate} & $0.1467$     & $0.3014$        & $0.5707$        & $0.1555$        & $0.3154$        & $0.5522$             \\
\hline
Case                        & \multicolumn{3}{c||}{Case 3, $R_{\mathrm{sum}}=1$}                    & \multicolumn{3}{c|}{Case 2, $R_{\mathrm{sum}}=2$}  \\
\cline{1-7}
User                        & {User 1}        & {User 2}        & {User 3}          & {User 1 (1 Layer)}        & {User 2 (2 Layers)}        & {User 3 (4 Layers)} \\
\cline{1-7}
Power                       & {$\frac{1}{7}$} & {$\frac{2}{7}$} & {$\frac{4}{7}$}  & {$\frac{1}{7}$} & {$\frac{2}{7}$} & {$\frac{4}{7}$}\\
\cline{1-7}
Path                        & \multicolumn{3}{c||}{$[1,1,1]\rightarrow[0.5,0.2,0.2]\rightarrow[0,0,0]$}  & \multicolumn{3}{c|}{\tabincell{c}{$[1,1,1]\rightarrow[0.9635, 0.7154, 0.0953]\rightarrow[0,0,0]$}}\\
\cline{1-7}
\tabincell{c}{Target rate}  & {$0.157$}       & {$0.281$}       & {$0.562$} & {$0.4$}        & {$0.7$}        & {$0.9$} \\
\cline{1-7}
{\tabincell{c}{Check edge\\distribution}} & $\eta_3=1$      & $\eta_4=1$      & $\eta_5=1$  & $\eta_4=1$ & $\eta_3=1$ & $\eta_3=1$\\
\cline{1-7}
{\tabincell{c}{variable degree\\set $\{d_{v}\}$}} & \multicolumn{3}{c||}{\{2:1:30, 35:5:50\}}  & \multicolumn{3}{c|}{\{2:1:50, 60:10:100\}}\\
\cline{1-7}
               &$\lambda_2,0.5250$    &$\lambda_2,0.3788$    & $\lambda_2,0.3289$   &$\lambda_2,0.4107$    &$\lambda_2,0.6158$    &$\lambda_2,0.5703$\\
Optimized &$\lambda_3,0.2138$    &$\lambda_3,0.1925$    & $\lambda_3,0.2277$   &$\lambda_3,0.2398$    &$\lambda_3,0.2261$    &$\lambda_3,0.1699$  \\
variable    &$\lambda_6,0.0978$    &$\lambda_6,0.0763$    & $\lambda_8,0.1189$   &$\lambda_8,0.1875$    &$\lambda_6,0.0726$    &$\lambda_6,0.1130$  \\
edge        &$\lambda_7,0.0618$    &$\lambda_7,0.1589$    & $\lambda_9,0.1747$   &$\lambda_9,0.0339$    &$\lambda_7,0.0413$    &$\lambda_7,0.0713$ \\
distribution &$\lambda_{30},0.0604$ &$\lambda_{50},0.1935$ & $\lambda_{45},0.1337$ &$\lambda_{10},0.0001$ &$\lambda_{44},0.0002$ &$\lambda_{16},0.0001$\\
                 &$\lambda_{35},0.0412$ &                      & $\lambda_{50},0.0161$ & $\lambda_{22},0.0001$ &$\lambda_{50},0.0440$ &$\lambda_{25},0.0277$\\
  &  &  &  &$\lambda_{35},0.0834$  &                      &$\lambda_{26},0.0477$ \\
  &  &  &  &$\lambda_{100},0.0445$ &                      &                       \\

\cline{1-7}
\tabincell{c}{Optimized\\rate} & $0.1588$        & $0.2926$        & $0.5608$          & $0.4145$             & $0.6844$        & $0.8652$  \\
\cline{1-7}
\end{tabular}
\end{table*}

As the ESE functions are readily available, according to the matching condition, we optimize the degree profile of LDPC codes  to match the ESE functions for user $k$, i.e.,
\begin{equation}\label{eq:DEC_transfer}
  v_k = \psi_{k}(\rho_k) = \left\{ \begin{array}{cc}
                         1, & \rho_k \leq \rho_{k,\min}, \\
                         \phi_{k}^{-1}(\rho_k), & \rho_{k,\min} \leq \rho_k \leq \rho_{k,\max}, \\
                         0, & \rho_k \geq \rho_{k,\max},
                       \end{array}
                       \right. \tag{23}
\end{equation}
where $\phi_{k}^{-1}(\rho_k)$ is the inverse of the ESE function of user $k$ and $\rho_{k,\min} = \frac{g_k}{\bm{g}^T\bm{1}-g_k+\sigma^2}$, $\rho_{k,\max} = \frac{g_k}{\sigma^2}$. Using the EXIT chart matching techniques~\cite{tenBrinkLDPCEXITTC04,Yuan2008,XYuanITI14,HanzoExitBICM}, the matching LDPC codes can be designed by properly choosing the degree distributions.

We basically follow the method described in \cite[Appendix 5G]{Yuan2008} to design irregular LDPC codes given a target transfer function $v = \psi(\rho)$, where $\rho$ is the \emph{a priori} SNR and $v$, the decoder output, denotes the extrinsic variance. The difference is that we use mutual information instead of the mean of LLR to track the evolution process.

The asymptotic performance of an LDPC code ensemble can be specified by its variable node and check node edge distribution polynomials, namely
\begin{equation}\label{eq:edge_distribution}
\lambda(x) = \sum\limits_{i=1}^{d_{v,\max}} \lambda_i x^{i-1} \mbox{~and~} \sum\limits_{i=1}^{d_{c,\max}} \eta_i x^{i-1}, \tag{24}
\end{equation}
where $\lambda_i$ (resp., $\eta_i$) is the fraction of edges in the bipartite graph of the LDPC code connected to variable nodes (resp., check nodes) with degree $i$, and $d_{v,\max}$ (resp., $d_{c,\max}$) is the maximum variable node (resp., check node) degree. Moreover, we use the Gaussian approximation~\cite{Chung2001}, i.e., \eqref{eq:GaussAssum1} and \eqref{eq:GaussAssum2},  to optimize the edge distributions for the sake of simplicity.

In~\cite{tenBrinkLDPCEXITTC04}, it is shown that the decoder characteristic for an LDPC code can be computed as
\begin{small}
\begin{align}
I_{E,V}\! &=\! \sum\limits_{\!i=1}^{\!d_{v,\max}}\!\lambda_i \cdot J \! \left( \sqrt{(i-1)\left[ J^{-1}\left(I_{E,C}\right) \right]^2 + 4\rho} \right),\tag{25} \label{eq:variable_evo} \\
I_{E,C}\! &=\! 1 -\! \sum\limits_{j=1}^{d_{c,\max}}\eta_j \cdot J\left( \sqrt{j-1} \cdot J^{-1}\left(1-I_{E,V} \right) \right), \tag{26} \label{eq:check_evo}
\end{align}
\end{small}
where $I_{E,V}$ (resp., $I_{E,C}$) is the extrinsic information from variable node (resp., check node) to check node (resp., variable node),  $\rho$ is the decoder input SNR, and the $J(\cdot)$ is defined by
\begin{equation}\label{eq:Jfunc}\nonumber
  J(\sigma_{ch}) = 1 - \int_{-\infty}^{\infty}\frac{e^\frac{-(y-\sigma_{ch}^2/2)^2}{2\sigma_{ch}^2}}{\sqrt{2\pi\sigma_{ch}^2}}\cdot \log_2\left[ 1+e^{-y} \right] dy,
\end{equation}
its inverse function is further denoted by $J^{-1}(\cdot)$. Substituting \eqref{eq:check_evo} into \eqref{eq:variable_evo}, we have \eqref{eq:evo}, where the LDPC code can be characterized by one single variable $I_{E,V}$.
\begin{figure*}
\begin{align}
     & \max\limits_{\{\lambda_i\}} \sum\limits_{i=1}^{d_{v,\max}} \frac{\lambda_i}{i} \nonumber \\
\mbox{s.t.~} & \sum\limits_{i=1}^{d_{v,\max}} \lambda_i = 1, \nonumber \\
     & \sum\limits_{i=1}^{d_{v,\max}}\!\lambda_i \cdot J \! \left( \sqrt{(i-1)\left[ J^{-1}\left( 1 -\! \sum\limits_{j=1}^{d_{c,\max}}\eta_j \cdot J\left( \sqrt{j-1} \cdot J^{-1}\left(1-I_{E,V} \right) \right) \right) \right]^2 + 4\rho} \right) > I_{E,V} \nonumber \\
     & \mbox{for~} \forall 0 < \rho < \infty \mbox{~and~} I_{E,V,\mbox{ini}}(\rho) \leq I_{E,V} \leq I_{E,V,\mbox{fin}}(\rho). \tag{28} \label{eq:LDPC_opt}
\end{align}
\hrulefill
\vspace*{-0.4cm}
\end{figure*}
The degree optimization problem can be formulated in (\ref{eq:LDPC_opt}). The cost function in (\ref{eq:LDPC_opt}) is to maximize the code rate. Let $I_{E,V,\mathrm{ini}}(\rho)$ be the initial extrinsic information given by the channel, which can be written as
\begin{align}
  I_{E,V,\mathrm{ini}}(\rho) &= \sum\limits_{\!i=1}^{\!d_{v,\max}}\!\lambda_i \cdot J \! \left( \sqrt{(i-1)\left[ J^{-1}\left(0\right) \right]^2 + 4\rho} \right) \nonumber \\
  &= J\left( 2\sqrt{\rho} \right). \tag{29}\label{eq:ini_IEV}
\end{align}
Let $I_{E,V,\mathrm{fin}}(\rho)$ denote the extrinsic information upon convergence, which corresponds to an output extrinsic variance to the MUD $v = \psi(\rho)$, i.e., given $\rho$, $I_{E,V,\mathrm{fin}}(\rho)$ should satisfy the following equation
\begin{align}\label{eq:output_variance}
  \sum_{i=1}^{d_{v,\max}} \Lambda_i \cdot f_{Q} \left( \frac{i \cdot \left[ J^{-1} \left( I_{E,C,\mathrm{fin}} \right) \right]^2}{4} \right) = \psi(\rho), \tag{30}
\end{align}
where $f_{Q}$ is defined in \eqref{eq:MMSEQPSK}, and
\begin{equation}
I_{E,C,\mathrm{fin}}\! = \! 1 -\! \sum\limits_{\!j=1}^{\!d_{\!c,\max}}\! \eta_j \cdot J\left( \sqrt{j-1} \cdot J^{-1}\left(1-I_{E,V,\mathrm{fin}} \right) \right) \nonumber
\end{equation}
is the converged message from check nodes to variable nodes. Furthermore, 
we  define
\begin{equation}
\Lambda_i = \frac{\lambda_i / i}{\sum\limits_{i=1}^{d_{v,\max}} \lambda_i / i} \nonumber
\end{equation}
as the fraction of variable node of degree $i$.  

\begin{algorithm}
\caption{Algorithm for LDPC Code Optimization in IDMA}
 \begin{algorithmic}[1]\label{alorithm:LDPCopt}
 \renewcommand{\algorithmicrequire}{\textbf{Input:}}
 \renewcommand{\algorithmicensure}{\textbf{Output:}}
 \REQUIRE Target decoder transfer functions $v_k = \psi_k (\rho_k )$, check edge distribution $\eta_k (x)$, maximum trial $T$, threshold $\epsilon$ and maximum variable degree $d_{v,\max}$.
 \ENSURE The optimized variable edge distribution $\lambda^{(T)}(x)$.
  \STATE Initialize $\lambda^{(0)}(x) = x$.
  \FOR {$t = 1$ to $T$}
  \STATE Solve (\ref{eq:LDPC_opt}) by linear programming to obtain $\lambda^{(t)}(x)$, where $I_{E,V,\mbox{fin}}(\rho)$ in (\ref{eq:LDPC_opt}) is obtained by solving (\ref{eq:output_variance}) using $\lambda^{(t-1)}(x)$.
  \IF {1-$\frac{\sum\limits_{i=1}^{d_{v,\max}}\lambda_i^{(t)}\lambda_i^{(t-1)}}{\sqrt{\left(\sum\limits_{i=1}^{d_{v,\max}}(\lambda_i^{(t)})^2\right) \left(\sum\limits_{i=1}^{d_{v,\max}}(\lambda_i^{(t-1)})^2\right)}} \leq \epsilon$}
  \STATE $\lambda^{(T)}(x) = \lambda^{(t)}(x)$.
  \RETURN $\lambda^{(T)}(x)$.
  \ENDIF
  \ENDFOR
 \RETURN $\lambda^{(T)}(x)$.
 \end{algorithmic}
\end{algorithm}
The optimization in (\ref{eq:LDPC_opt}) is a non-convex optimization. However, given $\{\Lambda_i\}$ and $\eta(x)$, the problem in (\ref{eq:LDPC_opt}) can be solved using standard linear programming. We use an iterative way to optimize the edge distribution $\lambda(x)$ with fixed $\eta(x)$ in Algorithm \ref{alorithm:LDPCopt}. In practice, Algorithm \ref{alorithm:LDPCopt} is repeated for several check edge distributions $\eta(x)$ till a matching code is found.

\subsection{Numerical Results}\label{sec:BER}
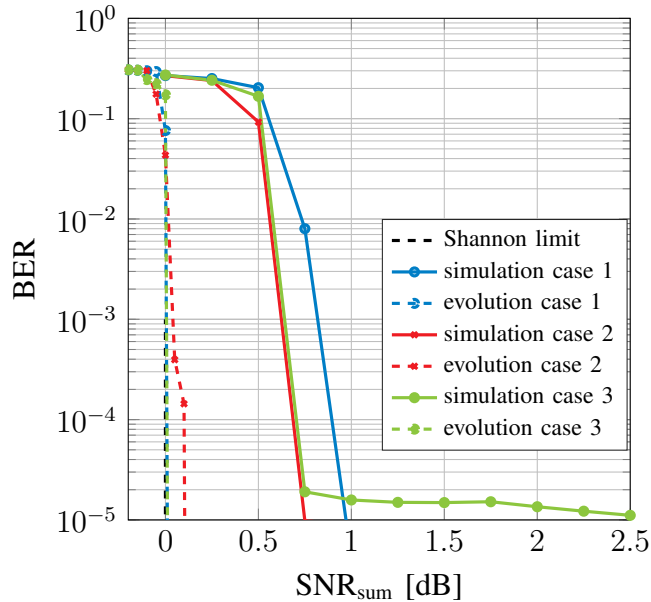
\begin{figure}[htbp]
    \centering
    \begin{tikzpicture}

\pgfplotsset{compat=1.12}
\begin{axis}[
        width=.5\linewidth,
	    height=.5\linewidth,
        xmajorgrids,
        yminorticks=true,
        ymajorgrids,
        yminorgrids,
        legend pos=south west,        
        legend style={at={(0.75,0.6)},
      anchor=north, legend columns=1, legend cell align=left,align=left,draw=white!15!black, font=\footnotesize},
        xlabel={$\textrm{SNR}_{\textrm{sum}}$ [dB]},
        ylabel={BER},
        ymode=log,
        mark size=1.5pt,
        xmin=-0.2,
        xmax=2.5,
        ymin=1e-5,
        ymax=1
    ]	
\addplot[color= black, very thick, dashed] coordinates {(0.0, 1e-10) (0.0,1e-3)};
\addlegendentry{Shannon limit}

      \addplot[color=mittelblau,mark=o, very thick] table {tikz/data/SNR_BER_5.dat};
	\addlegendentry{simulation case 1}
      \addplot[color=mittelblau,mark=o, very thick,dashed] table {tikz/data/SNR_BER_6.dat};
	\addlegendentry{evolution case 1}
	\addplot[color= rot, mark=x, very thick] table {tikz/data/SNR_BER_3.dat};
	\addlegendentry{simulation case 2}
      \addplot[color= rot,mark=x, very thick,dashed] table {tikz/data/SNR_BER_4.dat};
	\addlegendentry{evolution case 2}
	\addplot[color=apfelgruen, mark=*,very thick] table {tikz/data/SNR_BER_1.dat};
	\addlegendentry{simulation case 3}
	\addplot[color=apfelgruen, mark=*, very thick,dashed] table {tikz/data/SNR_BER_2.dat};
	\addlegendentry{evolution case 3}

\end{axis}

\end{tikzpicture}
    \vspace*{-0.4cm}
    \caption{BER curves and density evolution results for a three user MAC with matched codes; three different paths and QPSK signaling are considered.}
    \label{Fig:simulation}
    \vspace*{-1cm}
\end{figure}

With Algorithm~\ref{alorithm:LDPCopt}, if we use all variable degrees less than $d_{v,\max}$, the optimization can be quite slow. However, the optimized degree sequences are mostly comprised of a few small degrees. Therefore, we only use a subset of degrees less than $d_{v,\max}$ to run Algorithm~\ref{alorithm:LDPCopt} more efficiently. In the algorithm, we set $T=100$ and $\epsilon = 0.001$ for all optimizations. The optimized results as well as other parameters are summarized in Table~\ref{tab:OptResult}. Fig.~\ref{Fig:ESE_DEC_transfer} shows the optimized LDPC DEC transfer functions (denoted by dashed lines). The DEC functions match enough well with the ESE functions.

After matching the degree distribution, we construct parity check matrices for BER simulations. The parity-check matrix of code-word length $10^5$ for each user in each case  is randomly generated and subsequently we remove the cycle-$4$ loops in the matrix by edge permutation~\cite{McGowan2003}. Furthermore, we consider QPSK signaling in the simulation to verify that our Theorems work also well with finite alphabets, not only with Gaussian alphabets.

Fig.~\ref{Fig:simulation} shows the average BER performance for three users with matching codes along different decoding paths (denoted by solid lines, respectively), where we set a maximum iteration of $1000$ between the ESE detector and the LDPC decoders. Moreover, the Shannon limit at the sum-rate $R_{\textrm{sum}}=1$ along with the density evolution performance with optimized codes considering QPSK are provided. From the numerical results, we can conclude that
\begin{itemize}
  \item the evolution thresholds for three cases with QPSK are close to the Gaussian capacity. The loss incurred by finite alphabets is negligible at the target sum-rate.
  \item Three cases have BER below $10^{-4}$ within $1$ dB to the Shannon limit. The sum-rate capacity can be achieved for different paths also with QPSK signaling.
  \item For case 2, we computed a dedicated path to achieve an arbitrarily chosen rate tuple $(R_1, R_2, R_3) = (0.15, 0.3, 0.55)$. The numerical results also verified our path construction based on GA.
\end{itemize}

\begin{figure*}[htbp]
    \centering
   \begin{tikzpicture}


\begin{groupplot}[
    group style ={group size = 2 by 2, horizontal sep=1.5cm},
    width=0.45\linewidth,
    height=0.45\linewidth,
    xmajorgrids,
        ymajorgrids,
        zmajorgrids,
]
       \nextgroupplot [
         view={-37.5}{30},
        mark size=1.5pt,
         xmin=0,
        xmax=1,
        ymin=0,
        ymax=1,
        zmin=0,
        zmax=1,
        xlabel={$v_1$},
	ylabel={$v_2$},
	zlabel={$v_3$},
       ]
      \addplot3 [color=mittelblau, mark=o, very thick] table {tikz/data/DEC_Path6.dat};
      \addplot3 [color=mittelblau, mark=o,very thick,dashed] table {tikz/data/DEC_Path5.dat};
      \addplot3 [color=rot, mark=*,very thick] table {tikz/data/DEC_Path4.dat};
      \addplot3 [color=rot, mark=*,very thick, dashed] table {tikz/data/DEC_Path3.dat};
      \addplot3 [color=apfelgruen,mark=square,very thick] table {tikz/data/DEC_Path2.dat};
	\addplot3 [color=apfelgruen,mark=square,very thick,dashed] table {tikz/data/DEC_Path1.dat};
	
	\coordinate (top) at (rel axis cs:0,1);

     \nextgroupplot [
         view={0}{90},
        mark size=1.5pt,
         xmin=0,
        xmax=1,
        ymin=0,
        ymax=1,
        zmin=0,
        zmax=1,
        xlabel={$v_1$},
	ylabel={$v_2$},
	zlabel={$v_3$},
      legend to name={CommonLegend},
      legend style={legend columns=2}
       ]

      \addplot3 [color=mittelblau, mark=o,very thick] table {tikz/data/DEC_Path6.dat};
      \addlegendentry{simulation case 1}
      \addplot3 [color=mittelblau, mark=o,very thick,dashed] table {tikz/data/DEC_Path5.dat};
      \addlegendentry{evolution case 1}
      \addplot3 [color=rot, mark=*,very thick] table {tikz/data/DEC_Path4.dat};
     \addlegendentry{simulation case 2}
      \addplot3 [color=rot, mark=*,very thick, dashed] table {tikz/data/DEC_Path3.dat};
      \addlegendentry{evolution case 2}
      \addplot3 [color=apfelgruen,mark=square,very thick] table {tikz/data/DEC_Path2.dat};
      \addlegendentry{simulation case 3}
	\addplot3 [color=apfelgruen,mark=square,very thick,dashed] table {tikz/data/DEC_Path1.dat};
	\addlegendentry{evolution case 3}

	\coordinate (bot) at (rel axis cs:1,1);

\end{groupplot}

 \path (top)--(bot) coordinate[midway] (group center);
  \node[above, ,xshift=-20ex, yshift=18ex, right=1em,inner sep=0pt] at(group center -| current bounding box.north) {\pgfplotslegendfromname{CommonLegend}};

\end{tikzpicture}
    \caption{Evolution and simulation trajectories for three cases at $\mathrm{SNR_{sum}} = 1$~dB;
    Left: $3$-D diagram for a trajectory $(v_1, v_2, v_3)$;
    Right: Side view $(v_1, v_2)$ of a trajectory.
    }
    \label{Fig:trajectory}
\end{figure*}
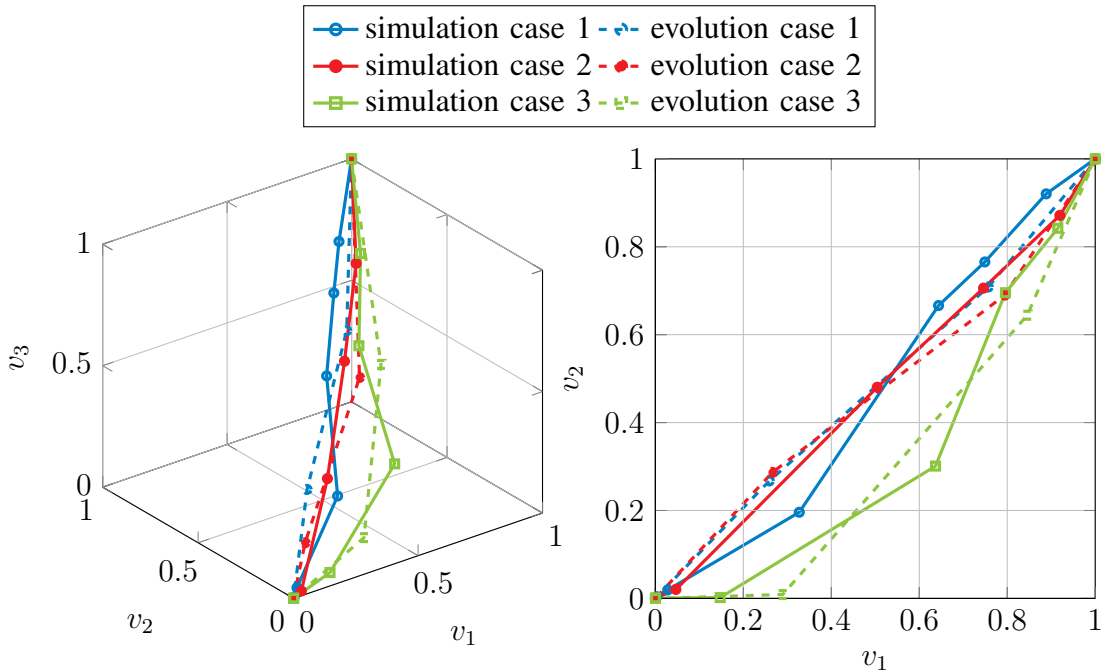
To further verify that the decoding path (or decoding trajectory $L(t)$) in the BER simulation is close to the desired path in the theory. We compare the decoding trajectories obtained via density evolution and BER simulation for three cases at the SNR $\mathrm{SNR_{sum}} = 1$~dB, where all users can decode its signal with high probability. The evolution trajectories differ from the specified paths (discussed in Sec.~\ref{sec:ESE-function}) mainly due to the different SNRs (the specified paths assume $\mathrm{SNR_{sum}} = 0$~dB). We observe that the simulation trajectories are consistent with those of density evolution for all three cases. This further consolidates the path independence theorem, and provides numerical evidence for finite alphabet cases.

\subsection{High Rate}
For scenarios where high rates per user is required, superposition coded modulation (SCM) \cite{LPSCMJSAC09} can be applied.
We consider the case that the data layers in SCM are of the same power.
Suppose that we have a $K$-user system with power allocation $[g_1, g_2, \cdots, g_K]^T$.
The decoding path $\bm{x}_0 \rightarrow \bm{x}_1 \rightarrow \cdots \rightarrow \bm{x}_n$ is also specified.
Then, we can convert this unequal power system into an equivalent equal power system as follows.
\begin{itemize}
	\item Find a real number $g > 0$ such that $L_i = g_i/g$ is an integer for all $i$.
	\item Change each $K$-dimension point $\bm{x}_i$ into a point $\bm{x}'_i$ with dimension $L = \sum_{i=1}^{K}L_i$ as follows.
			\begin{equation*}\label{layer_split}
			\begin{split}
				& \bm{x}'_i = \\
		 &		\left[\underbrace{x_{i,1}, \cdots, x_{i,1}}_{L_1 \mbox{~copies}}, \underbrace{x_{i,2}, \cdots, x_{i,2}}_{L_2 \mbox{~copies}}, \cdots, \underbrace{x_{i,K}, \cdots, x_{i,K}}_{L_K \mbox{~copies}}\right].
                     \end{split}
			\end{equation*}
	\item Finally, we have an equal power system with $L$ virtual users each with power $g$ and the decoding path is
	      $\bm{x}'_0 \rightarrow \bm{x}'_1 \rightarrow \cdots \rightarrow \bm{x}'_n$,
		  where the original user $i$ is the superposition of the virtual users with indices from $1+\sum_{j=1}^{i-1}L_j$ to $\sum_{j=1}^{i}L_j$.
\end{itemize}
By substituting the modified power allocation and path into \eqref{eq:rate},
it is easy to find that the equal power system satisfies the requirement of the original $K$-user system.

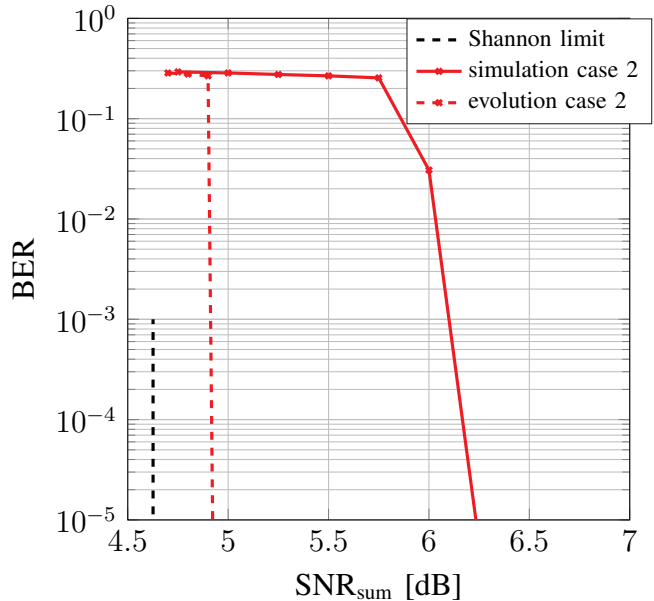
\begin{figure}
    \centering
	\begin{tikzpicture}

\pgfplotsset{compat=1.12}
\begin{axis}[
        width=.5\linewidth,
	    height=.5\linewidth,
        xmajorgrids,
        yminorticks=true,
        ymajorgrids,
        yminorgrids,
        legend pos=south west,        
        legend style={at={(0.8,1)},
      anchor=north, legend columns=1, legend cell align=left,align=left,draw=white!15!black, font=\footnotesize},
        xlabel={$\textrm{SNR}_{\textrm{sum}}$ [dB]},
        ylabel={BER},
        ymode=log,
        mark size=1.5pt,
        xmin=4.5,
        xmax=7,
        ymin=1e-5,
        ymax=1
    ]	
\addplot[color= black, very thick, dashed] coordinates {(4.626, 1e-10) (4.626,1e-3)};
\addlegendentry{Shannon limit}

	\addplot[color= rot, mark=x, very thick] table {tikz/data/HighRate3.dat};
	\addlegendentry{simulation case 2}
      \addplot[color= rot,mark=x, very thick,dashed] table {tikz/data/HighRate4.dat};
	\addlegendentry{evolution case 2}

\end{axis}

\end{tikzpicture}
    \caption{BER and density evolution results for a three-user MAC with matched codes at the sum-rate $R_{\mathrm{sum}}=1.964$; SCM and QPSK signaling are considered.}
    \label{Fig:simulation_high_rate}
\end{figure}

Table~\ref{tab:OptResult} shows the optimized LDPC code for a targeted sum-rate of $2$ for case 2, where the individual user rate is pre-defined and a dedicated decoding path is then specified to achieved that rate tuple.
For the simplicity of code design, we applied SCM to each user. In particular, the to be transmitted data packet for user 2 and user 3 is divided into two and four independent data layers. By doing this, the three user MAC system is converted to a MAC system with seven ``'users', each with the same transmit power.
The main motivation for applying SCM is that the curve matching code design becomes difficult at high rate \cite{ChulongCLIDMASCMA}.
It may require rather complicated joint design of modulation and coding scheme for each user. 
Simulation and density evolution results are depicted in Fig.~\ref{Fig:simulation_high_rate} for the most interesting case 2 (user rate is prescribed).
Compared to the low-rate scenario in Fig.~\ref{Fig:simulation}, the gap to Shannon-limit increases both for density evolution and simulation results at high rate (0.4~dB and 1.5~dB respectively). However, we conjecture that this gap can be reduced by imposing more data layers.

\section{Conclusion}\label{sec:conclusion}

It is proved under Gaussian approximation (GA) that the simple interleave-division multiple-access (IDMA), relying on a low-cost GA based multi-user detector (MUD), is capacity-achieving for general Gaussian multiple access channels
(GMAC) with arbitrary number of users, power distribution and with
single or multiple antennas. We show that IDMA with matching codes is capacity-achieving for arbitrary decoding path in the mean-square error (MSE) vector field. This property is further used to prove that IDMA achieves not
only the sum-rate capacity, but the entire GMAC capacity region. The
construction of capacity-achieving codes is also provided by establishing
an area theorem for multi-user extrinsic information transfer (EXIT)
chart. We provide numerical evidence supporting the GA and our achievable rate analysis.

\appendices{}

\section{Proof of \eqref{eq:UserRateGaussian}}\label{Proof1}

Let the SNR of ESE output $\rho_{k,\mathrm{min}},~\rho_{k,\mathrm{max}}$ as defined in \eqref{eq:SINRrange}.
Since the achievable rate formula in \eqref{eq:UserRateIntGeneral} requires the integration for the SNR $\rho$ spanning $\left(0,\infty\right)$, we explicitly write the transfer function of the DECs as 
\begin{align*}
v_{k} =
\begin{cases}
1, &\quad \rho\le\rho_{k,\mathrm{min}}\\
\psi_{k}\left(\rho_{k}\right),&\quad \rho_{k,\mathrm{max}}\le\rho\le\rho_{k,\mathrm{min}}\\
0 &\quad \rho\ge\rho_{k,\mathrm{max}}\\
\end{cases}
\end{align*} 
We also assume that the matching condition in \eqref{eq:MatchCondLine} holds.
Then, the achievable rates can be expressed as
\begin{align*}
R_{k} & ={\displaystyle \int_{\rho_{k,\mathrm{min}}}^{\rho_{k,\mathrm{max}}}}\frac{1}{\rho_{k}+v_{k}^{-1}}d\rho_{k}+{\displaystyle \int_{0}^{\rho_{k,\mathrm{min}}}}\frac{1}{\rho_{k}+1}d\rho_{k}
\end{align*}
Let $\rho'_{k}$ be the first derivative of $\rho_{k}$ with respect to $v_k$, we obtain
\begin{align*}
R_{k} & \overset{\rho'_{k}=\frac{d\rho_{k}}{dv_{k}}}{=}{\displaystyle \int_{v_{k}=1}^{v_{k}=0}}\frac{\rho'_{k}}{\rho_{k}+v_{k}^{-1}}dv_{k}+{\displaystyle \int_{0}^{\rho_{k,\mathrm{min}}}}\frac{1}{\rho_{k}+1}d\rho_{k}\\
 & ={\displaystyle \int_{1}^{0}}\frac{\rho'_{k}-v_{k}^{-2}+v_{k}^{-2}}{\rho_{k}+v_{k}^{-1}}dv_{k}+\underset{=w_{0}}{\underbrace{\mathrm{log}\left(1+\rho_{k,\mathrm{min}}\right)}}\\
 & =\left[\mathrm{log}\left(\rho_{k}+v_{k}^{-1}\right)+{\displaystyle \int}\frac{v_{k}^{-2}}{\rho_{k}+v_{k}^{-1}}dv_{k}\right]_{v_{k}=1}^{^{v_{k}=0}}+w_{0}\\
 & = \left[\mathrm{log}\left(\rho_{k}+v_{k}^{-1}\right)+{\displaystyle \int}\left(v_{k}^{-1}-\frac{1}{\rho_{k}^{-1}+v_{k}}\right)dv_{k}\right]_{v_{k}=1}^{^{v_{k}=0}}+w_{0}
\end{align*}
Let $g_{k}=P_{k}\left|h_{k}\right|^{2}$ be  the $k$th element of the vector $\mathbf{g}$ and $\boldsymbol{v}=\left[v_{1}, v_{2},\cdots,v_{K} \right]$,  we can express \eqref{eq:SINR} as $\rho_{k}=g_{k}/\left(\mathbf{g}^{T}\boldsymbol{v}-g_{k}v_{k}+\sigma^{2}\right)$ and obtain
\begin{align*}
R_{k} & \overset{(3\mathrm{a})}{=}\left[\mathrm{log}\left(\rho_{k}+v_{k}^{-1}\right)+\mathrm{log}\, v_{k} - {\displaystyle \int}\frac{g_{k}}{\mathbf{g}^{T}\boldsymbol{v}+\sigma^{2}}dv_{k}\right]_{v_{k}=1}^{^{v_{k}=0}} +w_{0}\\
 & =\left[\mathrm{log}\left(\rho_{k}v_{k}+1\right)-{\displaystyle \int}\frac{g_{k}}{\mathbf{g}^{T}\boldsymbol{v}+\sigma^{2}}dv_{k}\right]_{v_{k}=1}^{^{v_{k}=0}}+w_{0}\\
 & =-{\displaystyle \int}_{1}^{0}\frac{g_{k}}{\mathbf{g}^{T}\boldsymbol{v}+\sigma^{2}}dv_{k}
\end{align*}
where the last equality is due to $w_{0}=\mathrm{log}\left(1+\rho_{k,\mathrm{min}}\right) = \mathrm{log}\left(1+\rho_{k}\left(v_{k}=1\right)\right)$.

\section{Proof of Theorem \ref{thm:IDMA-achieves-everywhere}}\label{Proof2}

The user rate $R_{k}=-{\displaystyle \int}_{v_{k}=1}^{v_{k}=0}\frac{g_{k}}{\mathbf{g}^{T}\boldsymbol{v}+\sigma^{2}}dv_{k}$
is  a continuous and monotone decreasing function of $\boldsymbol{v}$.
If $v_{k}$ are unbounded, then $R_{k}$ can take on any value with the single
sum-rate constraint ${\displaystyle \sum R_{k}}\leq\mathrm{log}\left(\frac{\mathbf{g}^{T}\boldsymbol{1}+\sigma^{2}}{\sigma^{2}}\right)$.
In other words, there exists at least one integration path which allows achieving an arbitrary point within the region determined by ${\displaystyle \sum R_{k}}\leq\mathrm{log}\left(\frac{\mathbf{g}^{T}\boldsymbol{1}+\sigma^{2}}{\sigma^{2}}\right)$.
However, the value range of $R_{k}$ is constrained by the fact that
$0\leq v_{l}\leq1,\forall l$. 
Furthermore, the integrand $\frac{g_{k}}{\mathbf{g}^{T}\boldsymbol{v}+\sigma^{2}}$ is monotone decreasing with $v_l$, $\forall l \ne k$.
Therefore, the maximum of the $k$th user rate $R_k$ is attained when $v_{l}=0, \forall l\ne k$, i.e.,
\[
R_{k}\leq-{\displaystyle \int}_{v_{k}=1}^{v_{k}=0}\frac{g_{k}}{g_{k}v_{k}+\sigma^{2}}dv_{k}=\mathrm{log}\left(\frac{g_{k}+\sigma^{2}}{\sigma^{2}}\right).
\]
Similarly, the following constraints can also be obtained
\begin{align*}
R_{k}+R_{l} & \leq\mathrm{log}\left(\frac{g_{k}+g_{l}+\sigma^{2}}{\sigma^{2}}\right),\forall k\ne l\\
R_{k}+R_{l}+R_{m} & \leq\mathrm{log}\left(\frac{g_{k}+g_{l}+g_{m}+\sigma^{2}}{\sigma^{2}}\right),\forall k\ne l\ne m\\
\vdots & \vdots\\
{\displaystyle \sum R_{k}} & \leq\mathrm{log}\left(\frac{\mathbf{g}^{T}\boldsymbol{1}+\sigma^{2}}{\sigma^{2}}\right)
\end{align*}
and these constraints constitute the MAC capacity region. Hence, for any point in rate region determined by the above constraints (or equivalently the $K$-user MAC capacity region), there exists at least an integration path constrained by $v_l$, $\forall l \ne k$ achieving that point.

\balance

\bibliographystyle{IEEEtran}
\bibliography{bibliography}

\end{document}